\documentclass[traditabstract]{aa} % for the research notes
\usepackage{graphicx}
\usepackage{txfonts}
\usepackage{natbib}

\begin{document}
    \title{Flux- and volume-limited groups/clusters for the SDSS galaxies:\\
	catalogues and mass estimation\thanks{The group catalogue is available at {http://cosmodb.to.ee} and at the CDS via anonymous ftp to cdsarc.u-strasbg.fr (130.79.128.5).}}
	\titlerunning{Flux- and volume-limited groups/clusters for the SDSS}
    \author{E. Tempel\inst{1,}\inst{2}
	\and
	A. Tamm\inst{1}
	\and
	M.~Gramann\inst{1}
	\and
	T.~Tuvikene\inst{1,}\inst{3}
	\and
	L.~J. Liivam\"agi\inst{1,}\inst{4}
	\and
	I.~Suhhonenko\inst{1}
	\and
	R.~Kipper\inst{1,}\inst{4}
	\and
	M.~Einasto\inst{1}
	\and
	E. Saar\inst{1,}\inst{5}
    }

    \institute{Tartu Observatory, Observatooriumi~1, 61602 T\~oravere, Estonia\\
        \email{elmo.tempel@to.ee}
    \and
    National Institute of Chemical Physics and Biophysics, R\"avala pst 10, Tallinn 10143, Estonia
	\and
	Leibniz-Institut f\"ur Astrophysik Potsdam, An der Sternwarte 16, D-14482 Potsdam, Germany
	\and
	Institute of Physics, University of Tartu, 51010 Tartu, Estonia
	\and
	Estonian Academy of Sciences, Kohtu 6, Tallinn 10130, Estonia
    }

   %\date{Received December 20, 2011; accepted February 19, 2012}

  \abstract{We provide flux-limited and volume-limited galaxy group and cluster catalogues, based on the spectroscopic sample of the SDSS data release 10 galaxies. We used a modified friends-of-friends (FoF) method with a variable linking length in the transverse and radial directions to identify as many realistic groups as possible. The flux-limited catalogue incorporates galaxies down to $m_r=17.77~\mathrm{mag}$. It includes 588193 galaxies and 82458 groups. The volume-limited catalogues are complete for absolute magnitudes down to $M_{r,\mathrm{lim}} = -18.0$, $-18.5$, $-19.0$, $-19.5$, $-20.0$, $-20.5$, and $-21.0$; the completeness is achieved within different spatial volumes, respectively. Our analysis shows that flux-limited and volume-limited group samples are well compatible to each other, especially for the larger groups/clusters. Dynamical mass estimates, based on radial velocity dispersions and group extent in the sky, are added to the extracted groups. The catalogues can be accessed via \texttt{http://cosmodb.to.ee} and the Strasbourg Astronomical Data Centre (CDS).}


   \maketitle
%

%===================================================================== =====
\section{Introduction}

Galaxies, as well as stars or people, tend to gather in pairs, groups of several members, or even larger conglomerates. It is only natural that an individual surrounded by companions follows a somewhat different evolutionary path and develops a different appearance and inner qualities than a loner; the dependence of galaxy properties on the environment was shown already decades ago \citep{Einasto:74,Oemler:74,Davis:76,Dressler:80,Postman:84}. Therefore, catalogues of galaxy groups and clusters provide an unlimited data source for a wide range of astrophysical and cosmological applications, as illustrated below with just a random pick of the latest studies.

Among the most popular applications, galaxy group and cluster catalogues are used for quantifying the neighbourhood of galaxies in studies of the environmental dependencies of galaxy properties. The published catalogues have enabled a multitude of studies about the dependencies of the morphology, structure, gas content, star formation rate, merger rate, etc. of galaxies on their local environment \citep[some most recent examples include][]{Tempel:09, Carollo:13,Catinella:13,Hess:13,Kaviraj:13,Lackner:13,Peng:13}.

Besides, group and cluster catalogues can be used to distinguish central galaxies from their satellites \citep{Lacerna:13,Wetzel:13,Yang:13,Li:14}. They also simplify the selection and studying of the properties of specific galaxy systems, be it galaxy pairs, compact groups, loose groups, etc. Assuming that galaxy groups and clusters inhabit common dark matter haloes, catalogues can be used to probe dark matter haloes of different mass and to seek correlations between the mass and the galaxy content of the haloes \citep{Carollo:13,Hou:13,Huertas:13,McGee:13,Wetzel:13,Yang:13}, or to study the general group properties as a function of group mass \citep{Hearin:13, Budzynski:14} or as a function of the large-scale environment \citep{Luparello:13,Einasto:14}. By comparing the properties of observed groups and clusters to those produced in simulations, our understanding of structure formation can be validated \citep{Nurmi:13}.

Yet another broad field of applications for group and cluster catalogues is their usage for mapping the underlying cosmic web, e.g. by extracting large-scale filaments \citep{Zhang:13,Alpaslan:13}, or the largest known density enhancements, galaxy superclusters \citep{Zucca:93, Einasto:94, Einasto:01} from the spatial distribution of groups and clusters. Further, the properties of groups in superclusters, and galaxies in them can be studied \citep{Einasto:03a,Einasto:08, Einasto:11, Einasto:12, Lietzen:12, Krause:13}.
 
From the examples above, it is evident that catalogues of galaxy groups and clusters are in demand. But how to define a galaxy group and how to delineate it from galaxy redshift survey data? No straightforward recipe exists, different approaches would be valid for different science goals. In simulations we can see that gravitationally bound galaxy systems are linked together by an underlying dark matter halo, thus a good approach for defining a galaxy group or cluster would be through the existence of a common dark matter halo. However, for observational datasets, this method is of very little practical value. Instead, the friends-of-friends (FoF) algorithm has remained the most frequently applied means of identifying groups and clusters in galaxy redshift data ever since its introduction \citep{Turner:76,Press:82}. Today, many extensive FoF group and cluster catalogues are available, varying in the sample data and details of the group finder algorithm \citep[e.g. ][]{Tucker:00,Eke:04,Tago:06,Tago:08}.

The FoF method uses galaxy distances as the main basis of grouping, and is thus relatively simple and straightforward, while the membership of the produced groups is rather uncertain. Thus several other techniques for group and cluster extraction have been developed, well reviewed by \cite{Gal:06} and recently applied by \citet{Yang:05}, \citet{Koester:07}, \citet{Yang:07}, \citet{Hao:10}, \citet{Makarov:05}, \citet{Wen:12}, and \citet{MunosCuartas:12}. However, whether any method is more reliable than the others is still largely a matter of taste and debate \citep{Old:14}.

In \cite{Tempel:12a}, we constructed a flux-limited FoF group and cluster catalogue for galaxies in the Sloan Digital Sky Survey \citep[SDSS,][]{York:00} data release 8 \citep[DR8;][]{Aihara:11}. Here we present an update of this catalogue, based on SDSS~DR10 \citep{Ahn:13}. In addition to a flux-limited catalogue, we present volume-limited catalogues, valid within different galactic absolute luminosity bins (and thus being complete within different spatial volumes). We provide dynamical mass estimates for the detected galaxy systems, using the measured radial velocities and group extent in the sky. The catalogues also contain an rough estimate of the expected total luminosity of each group, assuming that some of the group members are not included on the sample due to observational limitations.

	Throughout this paper we assume the Wilkinson Microwave Anisotropy Probe (WMAP) cosmology: the Hubble constant $H_0 = 100\,h\ \mathrm{km\,s^{-1}Mpc^{-1}}$, the matter density $\Omega_\mathrm{m}=0.27$ and the dark energy density $\Omega_\Lambda=0.73$ \citep{Komatsu:11}.

%==========================================================================
\section{Data}

	\subsection{SDSS main sample}

\begin{figure}
	   \centering
	   \includegraphics[width=88mm]{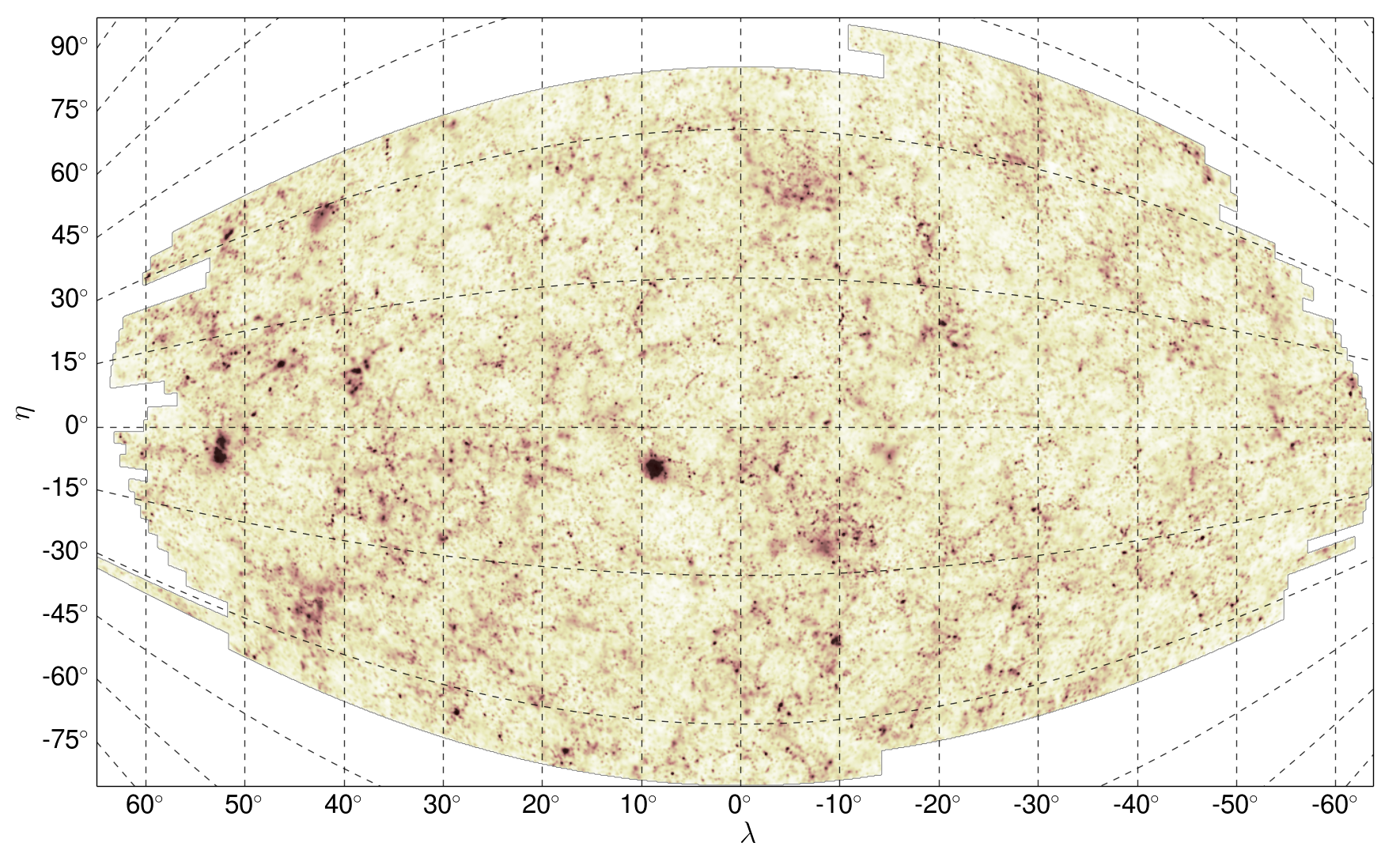}
	   \caption{SDSS main footprint in survey coordinates. Galaxy luminosity density field (integrated over the line of sight) is shown inside a sample mask as defined in \citet{Martinez:09}, yielding a sky coverage 7221 square degrees.}
	   \label{fig:footprint}
	\end{figure}

	The present work is based on the SDSS~DR10 \citep{York:00,Ahn:13}. We have utilised only the main contiguous area of the survey (the Legacy Survey). Since the survey edges in the sky are noisy in some regions, we applied the sample mask as defined by \citet{Martinez:09}. Figure~\ref{fig:footprint} shows the SDSS main footprint in the sky, covering 7221 square degrees (17.5\% from the full sky).

	Our previous galaxy group catalogue \citep{Tempel:12a} was based on the SDSS~DR8 \citep{Aihara:11}. While the sky coverage of the SDSS main area has remained unchanged already since DR7 \citep{Abazajian:09}, the data quality has improved. The SDSS webpage\footnote{\url{http://www.sdss3.org/dr10/imaging/caveats.php} \\ \url{http://www.sdss3.org/dr10/spectro/caveats.php}} lists several small caveats that have been corrected in DR9 \citep{Ahn:12} and DR10. Most notably, the astrometry of the objects has been corrected in DR9 and the imaging and spectroscopic pipeline have been updated/improved and applied to all images and spectra.

	The original data were downloaded from the Catalog Archive Server (CAS\footnote{\url{http://skyserver.sdss3.org/casjobs/}}) of the SDSS. For the primary selection, we used the \texttt{SpecObj} table as suggested by the SDSS team for spectroscopic objects. Our first selection included all objects with the spectroscopic class GALAXY or QSO; the final selection of the QSO objects was carried out manually later (see below). The corresponding photometric match was based on \texttt{bestobjid} (position-based match): for objects where it was undefined or the photometric class was not a GALAXY, it was based on \texttt{fluxobjid} (flux-based match). We only used those spectroscopic objects, for which the final matched photometric object class is GALAXY.
	
	The selected galaxy sample is cleaned and filtered by the SDSS team, however, due to the purely automatic filtering, some spurious entries still contaminate the sample. Most notable entries are bright over-saturated stars that are classified as galaxies. Since the luminosity of those objects is high, they would affect the bright-end of the galaxy luminosity function and the luminosity density field. Other spurious entries are large galaxies that sometimes have double/multiple entries in the sample. Also, for some of the galaxies, the given luminosity is wrong due to the proximity of bright stars and/or wrong sky estimation. In order to get rid of the spurious entires and to flag the galaxies where the luminosity is untrustworthy, we checked visually approximately 30\,000 galaxies using the SDSS Image List Tool\footnote{\url{http://skyserver.sdss3.org/dr10/en/tools/chart/listinfo.aspx}}. The steps that were taken to visually clean the sample are described below. In each step, we removed spurious entries and flagged galaxies where the luminosity was obviously incorrect. 
	We checked:
	\begin{itemize}
		\item  10\,000 apparently brightest galaxies (in $r$-band). For galaxies brighter than $m_r<13.5$ about 10\% of the objects were spurious. For galaxies $13.5<M_r<14.5$, about 1\% were spurious entries; this fraction decreases with luminosity;
		\item 5\,000 intrinsically brightest galaxies in the sample ($<1\%$ were spurious);
		\item 3\,000 intrinsically faintest galaxies in the sample (to ensure the correctness of the faint-end of the luminosity function);
		\item all the sources with the spectroscopic class QSO;
		\item all the objects with \texttt{bestobjid} missing or not GALAXY. For these objects, we used \texttt{fluxobjid} if the matched photometric object was classified as a galaxy;
		\item all the objects for which the difference between $r$-band point spread function (PSF) magnitude and model magnitude was smaller than 0.25 (thus further excluding some of the stellar sources in the catalogue);
		\item all the galaxies with the difference between $r$-band Petrosian and model magnitudes greater than 0.4;
		\item the entries where the colour indices $g-r$, $r-i$, and $g-i$ had extreme values;
		\item all the galaxy pairs that were closer than 5\arcmin\ and had roughly equal redshift (in order to remove double/multiple entries). This concerns mostly large nearby galaxies.
	\end{itemize} 
	Altogether about 600 entries were removed from the initial sample. In the final sample, the number of galaxies with incorrect luminosity (flagged entries in the final catalogue) is 1352.

	After the visual cleaning, we filtered the sample to contain only galaxies with the Galactic extinction corrected \citep[based on][]{Schlegel:98} Petrosian $r$-band magnitude $m_r\leq 17.77$. For fainter objects, the SDSS is incomplete \citep{Strauss:02}. After correcting the redshift relative to the motion with respect to the Cosmic Microwave Background (CMB), we set the upper distance limit at $z=0.2$.

	Our galaxy catalogue (see Appendix~\ref{app:cat}) includes all the relevant parameters from the SDSS CAS (apparent magnitudes, coordinates, observed redshift, etc). In order to facilitate the use of other SDSS parameters available in CAS, we include the \texttt{objid} and \texttt{specobjid} parameters in our catalogue. Appendix~\ref{app:cat} gives an example how to query additional parameters from CAS.

	In addition to the parameters provided by the SDSS team, we have added several parameters calculated/derived in this paper and explained in Appendix~\ref{app:cat}.

	\subsection{Spectroscopically complemented galaxy sample}

	The SDSS galaxy sample is not complete, mainly because of the fibre collision -- the minimum separation between spectroscopic fibres is 55\arcsec. For this reason, about 6\% of galaxies in the SDSS are without observed spectra. In addition, the redshift catalogue is incomplete for bright nearby objects due to the saturation limit of the SDSS detectors. \citet{Tempel:12a} studied the effect of missing galaxies on a group catalogue and concluded that the biggest impact is on galaxy pairs. In the SDSS sample, the absent galaxies are more likely to reside in groups and only 4\% of single galaxies have a companion missing. The estimated amount of missing pairs in the present group catalogue is about 8\%.

	In order to complement the SDSS spectroscopic sample, other redshift surveys covering the same sky area can be used. Following \citet{Choi:10}, we used the Two degree Field Galaxy Redshift Survey \citep[2dFGRS,][]{Colless:01,Colless:03}, the Two Micron All Sky Survey \citep[2MASS,][]{Jarrett:03,Skrutskie:06} Redshift Survey \citep[2MRS,][]{Huchra:12}, and the Third Reference Catalogue of Bright Galaxies \citep[RC3,][]{deVaucouleurs:91,Corwin:94} to complement the SDSS photometric objects with redshifts measured by one of the previously mentioned surveys. The addition of these objects improves the completeness for bright nearby objects and for objects in relatively nearby high-density regions (e.g. the Coma cluster); see \citet{Choi:10} for more details. All these three tables are listed in the SDSS CAS, and we downloaded the data from there.

	Matching of the SDSS photometric objects with these redshift catalogues was done using the sky coordinates. Only galaxies brighter than $m_r=17.77~\mathrm{mag}$ were considered. As before, we searched for duplicate entries. In addition, every matched galaxy was checked visually to avoid spurious entries.

	In total we added 3484 redshifts from the 2dFGRS dataset, 1119 redshifts from the 2MRS dataset and 280 redshifts from the RC3. Our final galaxy sample includes 588193 galaxies.

\begin{table*}
\caption{The SDSS main sample used for the flux- and volume-limited
	samples.}
\label{tab:vollim}
\centering
\begin{tabular}{lccccccccccc}
\hline\hline
Sample & $M_\mathrm{r,lim}$ & $z_\mathrm{lim}$ & $d_\mathrm{lim}$ & $N_\mathrm{gal}$ & $N_\mathrm{groups}$ &
$d_\mathrm{mean}$ & $LL$ & $b_\mathrm{gal}$ & in groups & \# den & weight \\
 & mag & & $h^{-1}\mathrm{Mpc}$ & & & $h^{-1}\mathrm{Mpc}$ & $h^{-1}\mathrm{Mpc}$ & & \% & $h^{3}\mathrm{Mpc}^{-3}$ &  \\
\hline
 \phantom{0}  & 1 & 2 & 3 & 4 & 5 & 6 & 7 & 8 & 9 & 10 & 11 \\
\hline
Flux-limited & 17.77\tablefootmark{a} & 0.200 & 574.2 & 588193 & 82458 & -- & 0.23--0.5 & -- & 47.9 & -- & -- \\

Vol-lim-18.0 & $-18.0$ & 0.045 & 135.0 & 49860  & 7328  & 3.294 & 0.380 & 0.115 & 59.8 & $4.11\cdot 10^{-3}$ & 1.135 \\
Vol-lim-18.5 & $-18.5$ & 0.057 & 168.9 & 73006  & 10929 & 3.638 & 0.411 & 0.113 & 58.3 & $3.11\cdot 10^{-3}$ & 1.204 \\
Vol-lim-19.0 & $-19.0$ & 0.071 & 211.0 & 105041 & 15715 & 4.027 & 0.445 & 0.110 & 55.7 & $2.29\cdot 10^{-3}$ & 1.315 \\
Vol-lim-19.5 & $-19.5$ & 0.089 & 261.3 & 149773 & 22524 & 4.433 & 0.480 & 0.108 & 53.8 & $1.73\cdot 10^{-3}$ & 1.504 \\
Vol-lim-20.0 & $-20.0$ & 0.110 & 322.6 & 163094 & 24258 & 5.321 & 0.515 & 0.097 & 47.9 & $9.88\cdot 10^{-4}$ & 1.856 \\
Vol-lim-20.5 & $-20.5$ & 0.136 & 397.2 & 164004 & 23007 & 6.541 & 0.548 & 0.084 & 39.9 & $5.01\cdot 10^{-4}$ & 2.596 \\
Vol-lim-21.0 & $-21.0$ & 0.168 & 486.2 & 125016 & 14155 & 8.766 & 0.577 & 0.066 & 28.1 & $1.68\cdot 10^{-4}$ & 4.463 \\
\hline
\end{tabular}
\tablefoot{
\tablefoottext{a}{Galactic-extinction-corrected Petrosian magnitude limit.}
Columns are as following.
\tablefoottext{1}{Absolute magnitude limit for volume-limited samples.}
\tablefoottext{2}{Maximum redshift.}
\tablefoottext{3}{Maximum comoving distance.}
\tablefoottext{4}{Number of galaxies in a sample.}
\tablefoottext{5}{Number of groups in a sample.}
\tablefoottext{6}{Mean pairwise separation of galaxies in comoving coordinates.}
\tablefoottext{7}{Used linking length (LL) in physical coordinates.}
\tablefoottext{8}{LL value in units of mean pairwise separation: $b_\mathrm{com}=b_\mathrm{gal}(1+z)$.}
\tablefoottext{9}{Fraction of galaxies in groups.}
\tablefoottext{10}{Group number density.}
\tablefoottext{11}{Luminosity weight for volume-limited samples.}
}
\end{table*}

\subsection{Volume-limited galaxy samples}
	\label{sect:data_vol}
	
	\begin{figure}
	   \centering
	   \includegraphics[width=88mm]{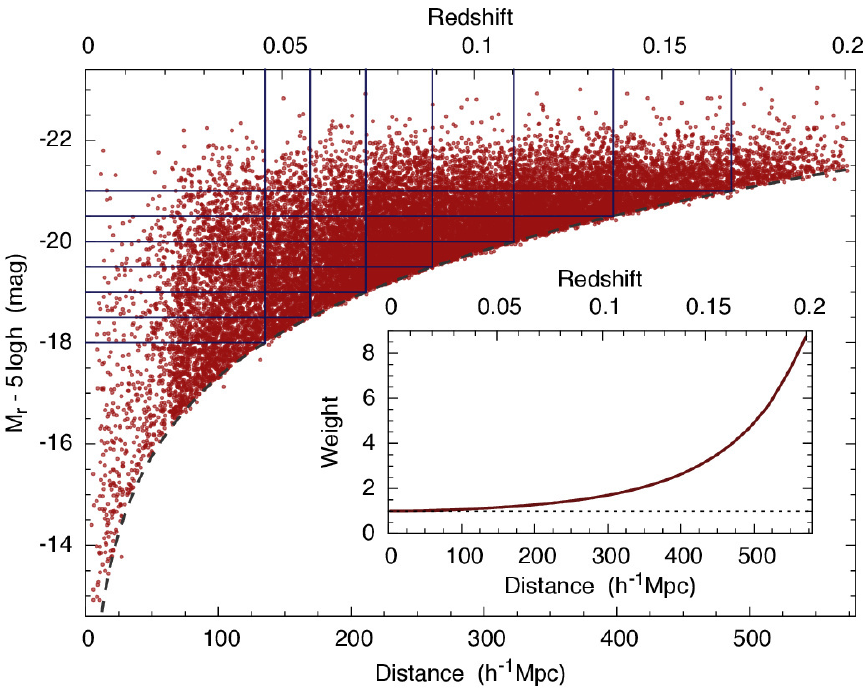}
	   \caption{Galaxy absolute magnitude in the $r$-filter as a function of comoving distance (lower axis) and redshift (upper axis). For visual clarity, only a small fraction of all galaxies is plotted: galaxies are randomly chosen to have approximately uniform number density with distance. Black dashed line shows the upper distance limit for a given absolute magnitude using the average $k$-correction. Solid blue lines show the subsamples for various volume-limited samples. Inner panel shows the luminosity weight factor as a function of distance. This weight takes into account the missing luminosity due to the flux-limited survey.}
	   \label{fig:dist_vs_flux}
	\end{figure}
	
	Intrinsically, the SDSS sample is flux-limited, meaning that the number density of galaxies decreases with distance: only the brightest galaxies are observed further away (see Fig.~\ref{fig:dist_vs_flux}). However, for many applications \citep[e.g. comparison with simulations,][]{Nurmi:13}, volume-limited samples are desired. Thus in addition to the flux-limited galaxy and group catalogue, we construct also volume-limited galaxy and group catalogues in this paper.

	To prepare the volume-limited galaxy sample, we started from the flux-limited galaxy and group samples (Sect.~\ref{sect:groups} gives details for our group finding method). Due to the peculiar velocities of galaxies in groups, the measured redshift (recession velocity) does not give an accurate distance to a galaxy located in a group/cluster \citep{Jackson:72}. Therefore, the apparent magnitude $m$ was transformed into the absolute magnitude $M$ (corresponding to redshift $z=0$) using the group/cluster mean redshift $z_\mathrm{cl}$ and mean comoving distance $d_\mathrm{com}$ (for isolated galaxies, we use the galaxy redshift and distance):
	\begin{equation}
		M_\lambda = m_\lambda - 25 -5 \log_{10}(d_L)-K_\lambda,
		\label{eq:absmag}
	\end{equation}
	where $d_L = d_\mathrm{com}(1+z_\mathrm{cl})$ is the luminosity distance in units $h^{-1}\mathrm{Mpc}$, $K_\lambda$ is the $k\,+\,e$-correction, and the index $\lambda$ refers to each of the $ugriz$ filters. The $k$-corrections were calculated with the KCORRECT (v4\_2) algorithm \citep{Blanton:07}. Evolution correction was estimated similarly by \citet{Blanton:03} assuming a distance-independent luminosity function. The estimation of the luminosity evolution is described in Appendix~\ref{app:lum_evol}.
	
	Figure~\ref{fig:dist_vs_flux} shows the absolute magnitude $M_r$ of the galaxies as a function of distance. We see that the faintest galaxies are missing further away. The inner panel in Fig.~\ref{fig:dist_vs_flux} displays the corresponding amount of luminosity that is missing at each distance (see Appendix~\ref{app:den} for details).

	We constructed 7 volume-limited galaxy samples with different magnitude ($M_r$) cuts: $-18.0$, $-18.5$, $-19.0$, $-19.5$, $-20.0$, $-20.5$, and $-21.0$. Since we complemented the original SDSS dataset with redshift measurements from other surveys, no upper magnitude limit is applied. The corresponding samples are shown in Fig.~\ref{fig:dist_vs_flux} as blue solid lines. The number of galaxies in each sample and the completeness distance limits are given in Table~\ref{tab:vollim}. The upper distance limit for the flux-limited sample is shown with the dashed line in Fig.~\ref{fig:dist_vs_flux} and is calculated from Eq.~(\ref{eq:absmag}) using the average $k$-correction near the survey limit.
	
	Usually the observed distances (redshifts) of galaxies are used to calculate absolute luminosities and to construct volume-limited samples. Considering that groups and clusters are extended objects in the redshifts space \citep[the so-called finger of god effect,][]{Tully:78}, we have used the flux-limited group mean distances here to construct the volume-limited samples (for galaxies in flux-limited groups we use the group mean redshift instead of the redshift of each galaxy). However, this choice affects only a small fraction of galaxies close to the magnitude limits. For the current dataset, the number of affected galaxies is less than 100 for each volume-limited sample. Resulting from the same effect, the volume-limited samples are affected also close to the upper distance limit. In this case, we have also decided to use flux-limited group mean distances instead of galaxy redshifts. We note that regardless of the choice of distance measurement, the galaxy and constructed group samples are always incomplete close to the survey boundaries.

%==========================================================================
\section{Method}

\subsection{Flux-limited galaxy groups/clusters}
\label{sect:groups}

	In this study, we followed the basic steps of the method described in \citet{Tago:08,Tago:10} to extract groups from the flux-limited galaxy sample. Below, the method is briefly outlined and the improvements with respect to \citet{Tago:08,Tago:10} are given.

	Our group finding is based on the friend-of-friend (FoF) algorithm \citep{Turner:76,Press:82}. The FoF method links galaxies into systems, using a certain neighbourhood radius, the linking length (LL). For each galaxy, all neighbours within the LL radius are considered to belong to the same system. The number and richness of the detected groups strongly depend on the chosen LL. In most cases, LL is not taken constant, but is allowed to vary with distance and/or other parameters.

	Our experience shows that the choice of the LL depends on the goals of the specific study. Here, our aim is to find as many groups as possible, while keeping the general group properties uniform with respect to distance. In our group definition, we have tried to avoid the inclusion of large sections of the surrounding large-scale filaments and parts of superclusters.
	
	The applied FoF group construction method has been tested on mock galaxy surveys in \citet{Nurmi:13} and \citet{Old:14}. Both papers indicate that the used FoF parameters (LL in radial and transversal direction) provide statistically reliable groups. In addition, the FoF parameters used here are in good agreement with the values applied by others \citep[see][]{Duarte:14}.

\begin{figure}
	   \centering
	   \includegraphics[width=88mm]{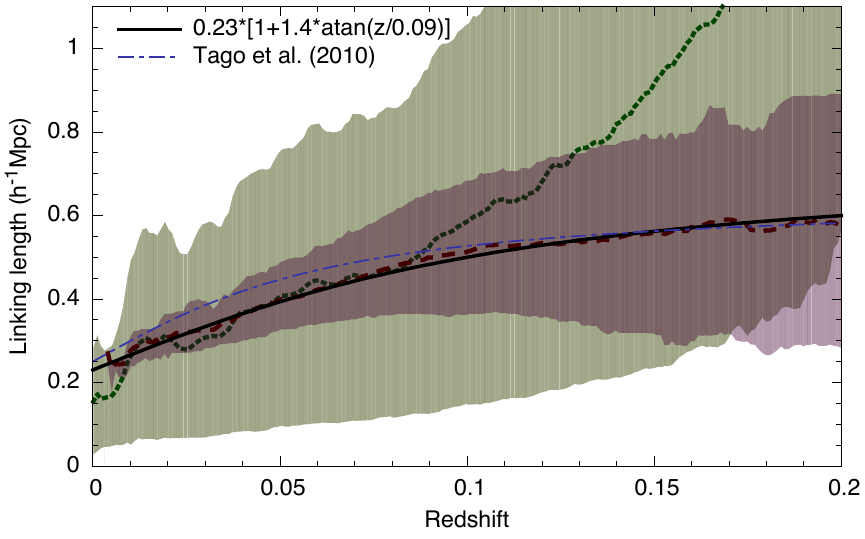}
	   \caption{FoF LL (in physical coordinates) as a function of redshift. Green dotted line shows the mean distance to the nearest neighbouring galaxy in physical coordinates, with 68\% confidence limits (green region). Red dashed line shows the LL scaling derived after shifting the nearby groups to larger distances, together with 68\% confidence limits (red region). Black solid line shows the best fit to the scaling relation. Blue dotted-dashed line shows the scaling from \citet{Tago:10}.}
	   \label{fig:linking_length}
	\end{figure}

	To find the proper scaling of LL with distance, we firstly calculated the mean distance (in physical coordinates) to the nearest galaxy in the plane of the sky. The neighbour was sought within a cylindrical volume: the ratio of the radial to the transversal LLs was taken 10 (after transforming the radial LL in units $\mathrm{km}\,\mathrm{s}^{-1}$ into a formal distance in $h^{-1}\mathrm{Mpc}$). The smallest cylinder that contains two galaxies defines the minimum distance between the two galaxies, and thus also the scaling for the LL. The cylinder diameter as a function of distance is shown in Fig.~\ref{fig:linking_length} with the green shaded area, the dotted green line shows the running mean of the distances. Below $z=0.1$, the nearest neighbour galaxy usually belongs to the same group. Further away, the nearest galaxy typically belongs to another group, therefore the distance between neighbouring galaxies is increasing rapidly. This increase of the mean distance is an expected effect of a flux-limited survey.
	A similar LL determination method is used by \citet{Old:14}, where various group finding algorithms (including the one presented here) are compared. \citet{Old:14} shows that the groups extracted using such LL strategy are statistically correct and the corresponding group properties are meaningful.

	\begin{figure}
	   \centering
	   \includegraphics[width=80mm]{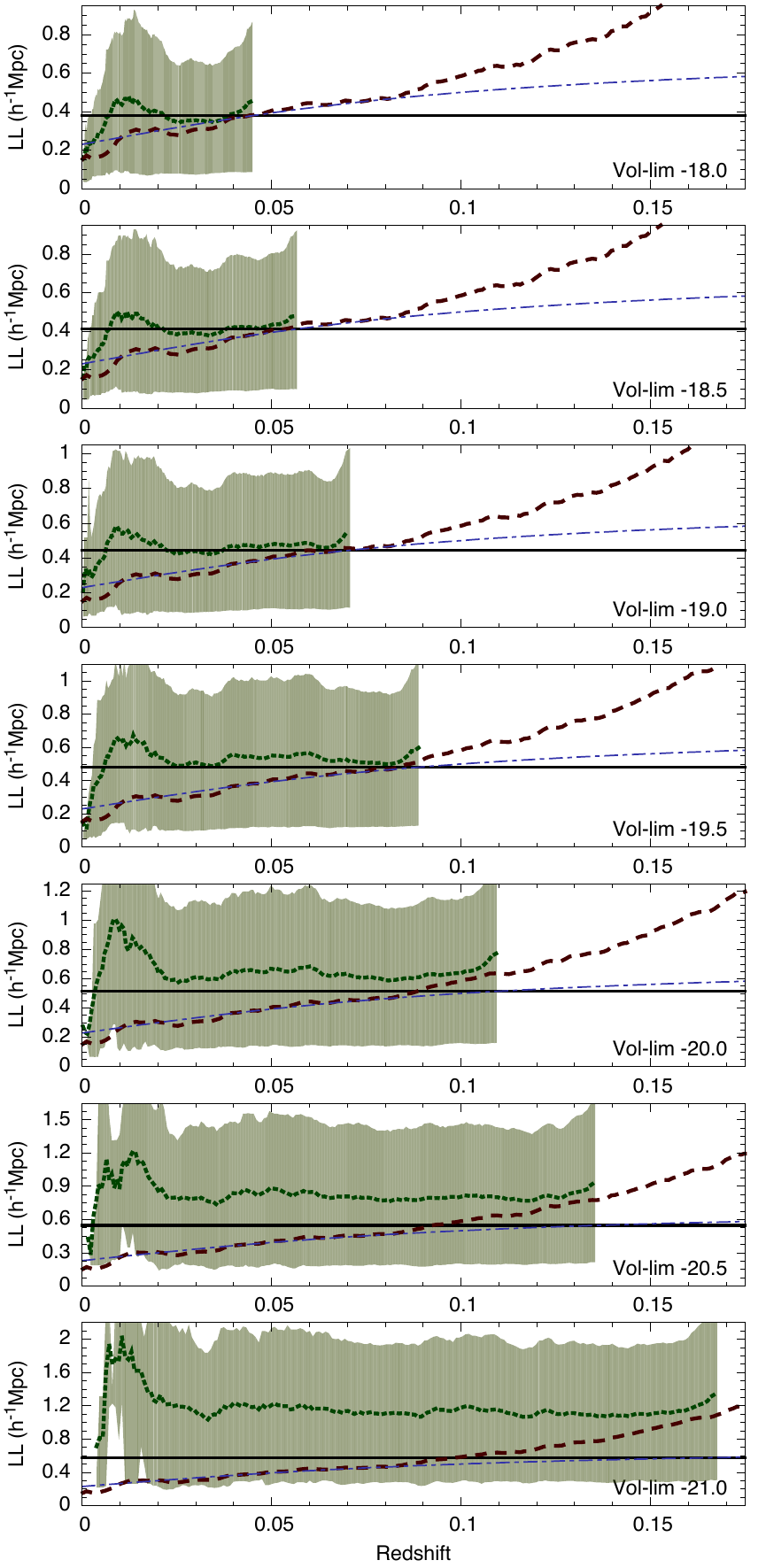}
	   \caption{Mean distance to the nearest galaxy (in physical coordinates) in various volume-limited samples together with 68\% confidence limits (green dotted lines with shaded areas) as a function of redshift. For comparison, mean distance to the nearest galaxy in  the flux-limited sample is shown (red dashed lines). Blue dashed-dotted lines show the LL scaling relation as found for flux-limited sample and black solid lines show the used LL for volume-limited samples (in physical coordinates).}
	   \label{fig:linking_length_vollim}
	\end{figure}
	
	To construct our flux-limited group sample, we made an initial guess for the LL. For that, an arctan law
	\begin{equation}
		d_{LL}(z) = d_{LL,0}\left[ 1+a\arctan(z/z_\star) \right]
		\label{eq:atan}
	\end{equation}
	was fitted to the scaling relation of the nearby galaxy sample ($z < 0.1$). In Eq.~(\ref{eq:atan}) $d_{LL,0}$ is LL at $z=0$, and $a$ and $z_\star$ are free parameters. Using the resultant LL law, we created a test group catalogue. Next we selected all the groups within the nearby volume ($d < 200~h^{-1}\mathrm{Mpc}$) with 15 to 50 members (altogether 539 groups). Smaller groups were excluded to reduce noise and the largest groups were expelled in order to avoid large nearby clusters. The remaining richness range represents well the groups in the nearby Universe. Naturally, the scaling relation depends on the used group richness bin, with smaller groups yielding smaller LL values, and vice versa. The range of 15--50 members was chosen here in order to roughly match with the mean distance to the nearest galaxy in the plane of the sky, e.g. to match with the initial scaling relation.

	Our next task was to find similar groups at distances, where only the brightest members are seen. Assuming that the group members are all at the mean distance of the group, we determined their absolute magnitudes and peculiar radial velocities. Then we re-calculated the parameters of the groups as they would appear if the groups were located at larger distances. As more and more fainter group members fall outside the observable apparent magnitude limit with increasing distance, the group membership changes. We calculated the minimum FoF LL necessary to keep the group together at each distance. Since by definition groups are gravitationally bound systems, we kept the physical size of the groups while shifting them to greater distances. The mean required LL value (in physical coordinates) is shown in Fig.~\ref{fig:linking_length} with the red dashed line. The figure shows that in order to keep the group finding consistent, the LL has to be increased moderately with distance as in our previous papers \citep{Tago:08,Tago:10,Tempel:12a}. We fitted an arctan law to the new scaling relation and created a new test group catalogue and repeated the procedure. After a few iterations, the relation converged at the following parameters: LL at a redshift $z=0$ is 0.23~$h^{-1}\mathrm{Mpc}$ and the parameters are $a=1.4$ and $z_\star = 0.09$. The final law is shown in the Fig.~\ref{fig:linking_length}, where we also show the scaling relation as found in \citet{Tago:10}.

	Since galaxy locations in the radial direction are drawn from the redshift space, galaxy groups appear elongated along the radial direction. We have taken the ratio of the radial LL (in $h^{-1}\mathrm{Mpc}$) to the transversal LL (in $\mathrm{km}\,\mathrm{s}^{-1}$) to be 10, yielding a initial radial LL value 230~$\mathrm{km}\,\mathrm{s}^{-1}$. The value 10 corresponds to the average elongation of the groups along the radial direction (see Fig.~\ref{fig:testgr_shape}). The resultant shape of the used LL distribution is cylindrical. One may argue whether it would be better to use ellipsoidal LL distribution. Tests have shown however that the cylindrical kernel reproduces recovered galaxy groups better \citep{Eke:04}.

	As shown in \citet{Tago:08,Tago:10}, the given LL determination leads to reasonable group properties. Our final group catalogues are rather homogeneous. The richness, mean size, and velocity dispersion of a group are practically independent of the group distance (see below). The homogeneity of our catalogues have been tested also by other authors. For example, \citet{Tovmassian:09} has selected poor groups from our previous SDSS catalogues and has concluded that the main parameters of our groups are distance-independent and well suited for statistical analysis.

	Our final group catalogue contains 82458 groups with two or more members. Almost half of the galaxies in the sample (48\%) belong to a group.
	
	We note that the choice of the LL in transversal and radial direction is rather arbitrary and we have chosen the values based on our previous experiences. In several papers \citep[e.g.][]{Eke:04,Berlind:06,Yang:07,Robotham:11} the mock catalogues are used to find the `best' values for FoF parameters. However, no universally good value exists. \citet{Duarte:14} analysed the effect of LL on the detected groups and conclude that the choice of the LL determines the fragmentation, merging, completeness, and reliability of the constructed group catalogue. They also show that the LL values used here are close to the average of the values applied by others. Therefore, according the current knowledge, the group finding algorithm presented in this paper stands on a solid ground.

\subsection{Construction of volume-limited groups/clusters}

	Volume-limited groups are based on volume-limited galaxy samples as described in Sect.~\ref{sect:data_vol}. Since the galaxy number density in volume-limited samples is constant by definition, we have used a constant LL (in physical coordinates) for each volume-limited sample. The LL was chosen to produce groups that are statistically similar to the ones found for flux-limited sample. To achieve this, the LL was chosen based on the LL scaling found for the flux-limited sample. For each volume-limited sample, the used LL is scaled according to the upper distance limit of the given volume-limited sample. At this distance, the number density of galaxies in the flux-limited sample is roughly the same as in the volume-limited sample. The used LL values are given in Table~\ref{tab:vollim} and shown in Fig.~\ref{fig:linking_length_vollim} as black solid lines.

	It is important to note that we intended to detect (mostly) virialised groups. We have therefore considered the LL value in the physical instead of the comoving coordinates, expecting the groups to resist the cosmological expansion. In our previous volume-limited catalogues \citep{Tago:10}, comoving coordinates were used. However, the difference between the two choices is relatively small within the given redshift range, so it does not affect the general properties of the produced groups.

	In general, there is no strict rule for choosing the LL for the FoF algorithm and the LL is slightly different in various papers. For example, \citet{Berlind:06} uses the value $b_\mathrm{com}=0.14$ and the ratio of the radial to the transversal LL was taken 5.4. The value $b_\mathrm{com}$ is given in units of the mean pairwise separation of galaxies in the survey. \citet{Eke:04} adopts the values $b_\mathrm{com}=0.13$ and the ratio of the radial to the transversal LLs was taken 11. The value ($b_\mathrm{com}=b_\mathrm{gal}(1+z)$) in our catalogue is redshift dependent (due to the constant LL in physical units) and is slightly lower (up to 0.12) than that used by \citet{Eke:04} and \citet{Berlind:06}. The ratio of the radial to the transversal LLs in our catalogues is slightly lower than in \citet{Eke:04} and higher than in \citet{Berlind:06}. Compared to our previous volume-limited catalogues based on SDSS DR7 \citep{Tago:10}, the LL values used in this paper are larger by up to 50\%, depending on the volume-limited sample and distance.
	
	Recently, \citet{Duarte:14} performed an in-depth study for the choice of the LL
in FoF algorithms. They conclude that the optimal LL depends on the scientific goal for the group catalogue. The LL values used in our catalogue are within the range proposed by \citet{Duarte:14}.
	
	Figure~\ref{fig:linking_length_vollim} shows the mean distance to the nearest galaxy (in physical coordinates) in the volume-limited samples as a function of distance (green dotted lines). For comparison, the mean distance in the flux-limited sample is shown (red dashed lines). Figure~\ref{fig:linking_length_vollim} shows that in the volume-limited samples, the mean distance to the nearest galaxy is almost independent of redshift as expected. The deviations in the nearby region ($z<0.02$) are caused by the small sample volume. The slight increase close to the upper distance limit is due to the limit itself.
	
	For a comparison, Fig.~\ref{fig:linking_length_vollim} shows the LL scaling relation for the flux-limited sample (blue dotted-dashed lines) and the used LL in the volume-limited samples (black solid lines). We note that for the four fainter volume-limited samples, the used LL is roughly the same as the mean distance to nearest galaxy. For the brighter volume-limited samples, the mean distance to the nearest galaxy is larger than the used LL. This is because the used LL scaling relation in the flux-limited sample deviates from the mean distance to the nearest galaxy for redshifts $z>0.1$. Since our aim was to construct volume-limited group catalogues that are comparable with the flux-limited catalogue, the LL for the volume-limited samples is taken from the scaling relation found for the flux-limited sample, rather than found independently for each volume-limited sample. This choice gives us volume-limited groups that are well comparable with the flux-limited groups (see Sect.~\ref{sect:res_vollim}).

\begin{figure}
	   \centering
	   \includegraphics[width=88mm]{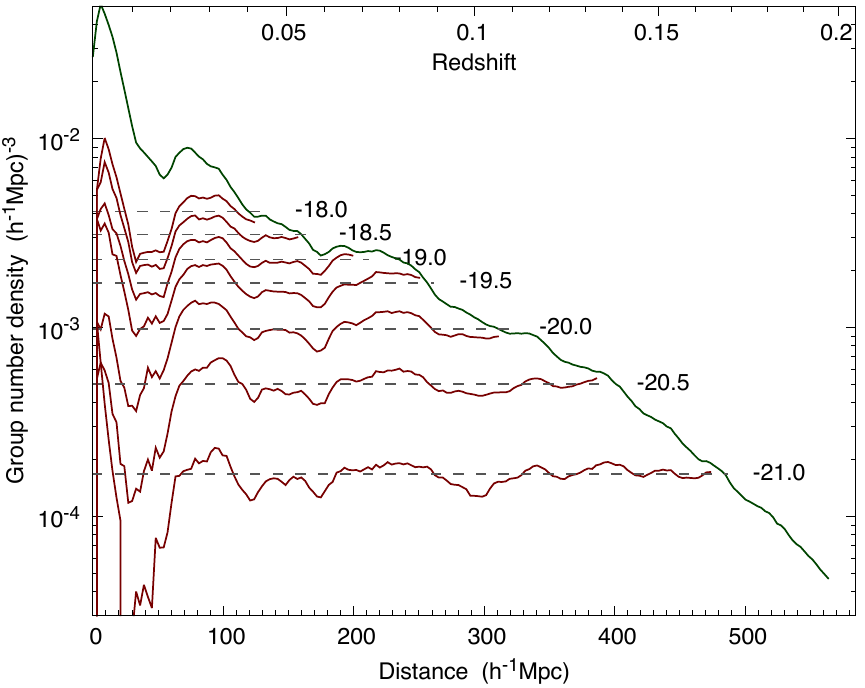}
	   \caption{Group number density as a function of distance for flux- and volume-limited samples. Green line shows the flux-limited sample, red lines show different volume-limited samples. Dashed lines give the average number density for the volume-limited samples (the values are given in Table~\ref{tab:vollim}). The density minimum at lower distances is caused by the low statistics and a small volume.}
	   \label{fig:gr_nr_dens}
	\end{figure}

Figure~\ref{fig:gr_nr_dens} shows the number density of groups for the flux-limited sample and for various volume-limited samples. The number density of groups in the flux-limited sample is decreasing. For the volume-limited samples it is almost constant as expected. The only differences are seen in the nearby regions, where the cosmic variance due to the small volume has a large effect. This figure shows that the groups extracted from the volume-limited samples are statistically homogeneous across the given volumes. Note that Fig.~\ref{fig:gr_nr_dens} does not characterise the group finding method itself. The deviations seen in the distributions are caused by the large-scale structure and the limited sample sizes.

\subsection{Group and galaxy properties}
\label{sect:group_prop}

In this section, we describe the galaxy and group parameters that are included in the catalogues. These were given also in our previous catalogue \citep{Tempel:12a}.

For each galaxy, we have estimated the galaxy morphology as described in \citet{Tempel:11a}. In \citet{Tempel:12a} the morphologies were compared to the ones provided by \citet{Huertas-Company:11} and a very good agreement was found. In the present catalogue, galaxies are classified as spirals or ellipticals, if the morphology in both catalogues agree. Otherwise the morphology is marked unclear.

    The  velocity dispersion $\sigma_v^2$ for groups were calculated with the standard formula
\begin{equation}
    \sigma_v^2 = \frac{1}{(1+z_\mathrm{m})^2(n-1)}\sum\limits^{n}_{i=1}(v_i-v_\mathrm{mean})^2,
	\label{eq:sigv}
\end{equation}
where $v_\mathrm{mean}$ and $z_\mathrm{m}$ are the mean group velocity and redshift, respectively, $v_i$ is the velocity of an individual group member, and $n$ is the number of galaxies with observed velocities within the group.

	\begin{figure}
	   \centering
	   \includegraphics[width=88mm]{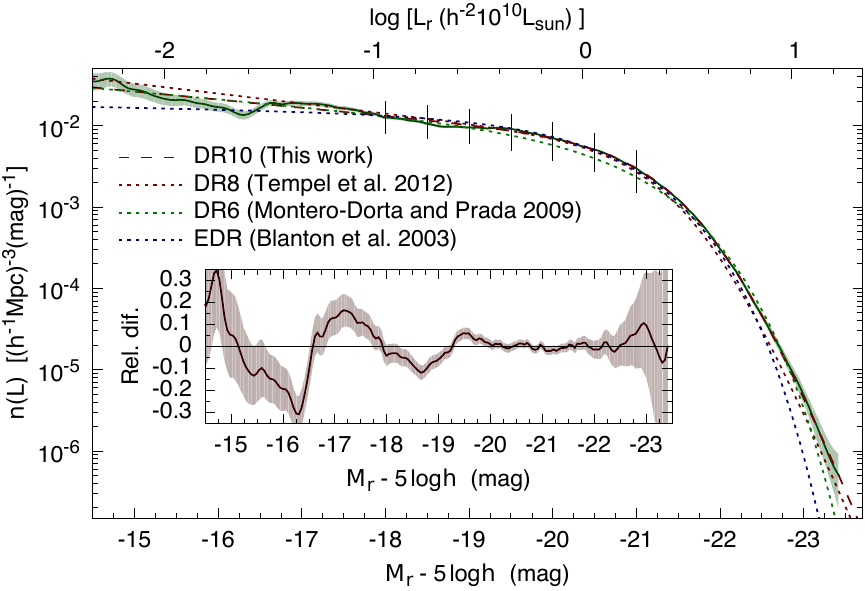}
	   \caption{Galaxy luminosity function in the $r$-filter. The solid line and the shaded region show the measured luminosity function together with the 99.7\% confidence limit (corresponds to $3\sigma$). Dashed line shows the best-fit double-power law fit. Dotted lines show previous approximations by \citet[][EDR]{Blanton:03},  \citet[][DR6]{Montero-Dorta:09}, and \citet[][DR8]{Tempel:12a}. The latter two are approximations with Schechter functions. Inner panel shows the relative difference between the best-fit double-power law function and the measured luminosity function. Vertical lines show the magnitude limits for the volume-limited samples.}
	   \label{fig:lumfun}
	\end{figure}

    The extent of the group in the plane of the sky is defined as
\begin{equation}
    \sigma_\mathrm{sky}^2 = \frac{1}{2n(1+z_\mathrm{m})^2}\sum\limits^{n}_{i=1}(r_i)^2,
	\label{eq:sigr}
\end{equation}
where $r_i$ is the projected distance in the sky from the group centre (in comoving coordinates, in units of $h^{-1}$Mpc), and $z_\mathrm{m}$ is the mean redshift of the group.

	The virial radii $R_\mathrm{vir}$ of the groups are calculated from the formula
	\begin{equation}
		\frac{1}{R_\mathrm{vir}} = \frac{2}{(1+z_\mathrm{m})n(n-1)}\sum\limits^{n}_{i\neq j}\frac{1}{R_{ij}},
	\end{equation}
	where $R_{ij}$ is the projected distance between galaxies in pairs in a group.

	\begin{figure}
	   \centering
	   \includegraphics[width=88mm]{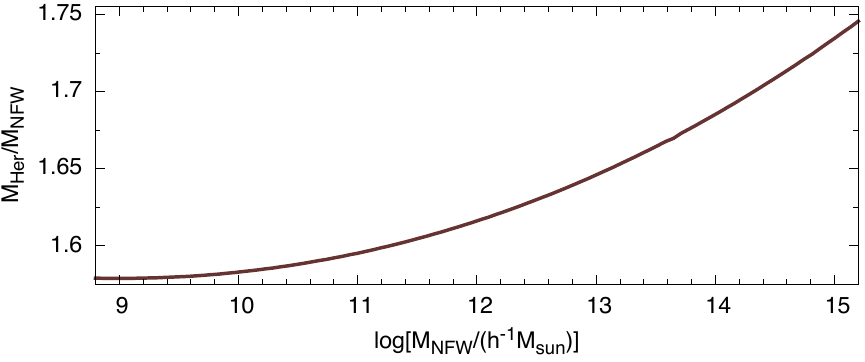}
	   \caption{The ratio of group dynamical masses calculated assuming Hernquist and NFW profiles as a function of group mass.}
	   \label{fig:gr_mass_nfw_her}
	\end{figure}
	
	Figure~\ref{fig:lumfun} shows the luminosity function in the $r$-filter for our final sample of the SDSS galaxies. It is calculated using an adaptive smoothing kernel as described in \citet{Tempel:11a}. The vertical lines mark the limits for the different volume-limited samples, described above. For comparison, we show the best-fit analytical luminosity functions from previous data releases: \citet[][DR8]{Tempel:12a}, \citet[][DR6]{Montero-Dorta:09}, and \citet[][EDR]{Blanton:03}.

We have used the double-power-law to find an analytical approximation of the luminosity function (see Fig.~\ref{fig:lumfun})
\begin{equation}
        n (L) \mathrm{d}L \propto 
		(L/L^{*})^\alpha \left[1 + (L/L^{*})^\gamma\right]^\frac{\delta-\alpha}{\gamma} \mathrm{d}(L/L^{*}), 
        \label{eq:abell}
\end{equation}
where $\alpha$ is the exponent at lower luminosities $(L/L^{*}) \ll 1$, $\delta$ is the exponent at higher luminosities $(L/L^{*}) \gg 1$, $\gamma$ is a parameter that determines the speed of transition between the two power laws, and $L^{*}$ is the characteristic luminosity of the transition. We find the best match with the observed luminosity function with $\alpha=-1.250\pm0.008$, $\delta=-7.32\pm0.30$, $\gamma=1.71\pm0.05$, and $M^{*}=-21.88\pm0.06$ (corresponds to $L^{*}$).

Additionally, the catalogues include environmental densities of the galaxies and groups. These densities are important when analysing the influence of the local and/or global environments on galaxy evolution. Local and global densities are estimated using different smoothing radii. The density calculation is described in Appendix~\ref{app:den}.

\begin{figure*}
   \centering
   \includegraphics[width=180mm]{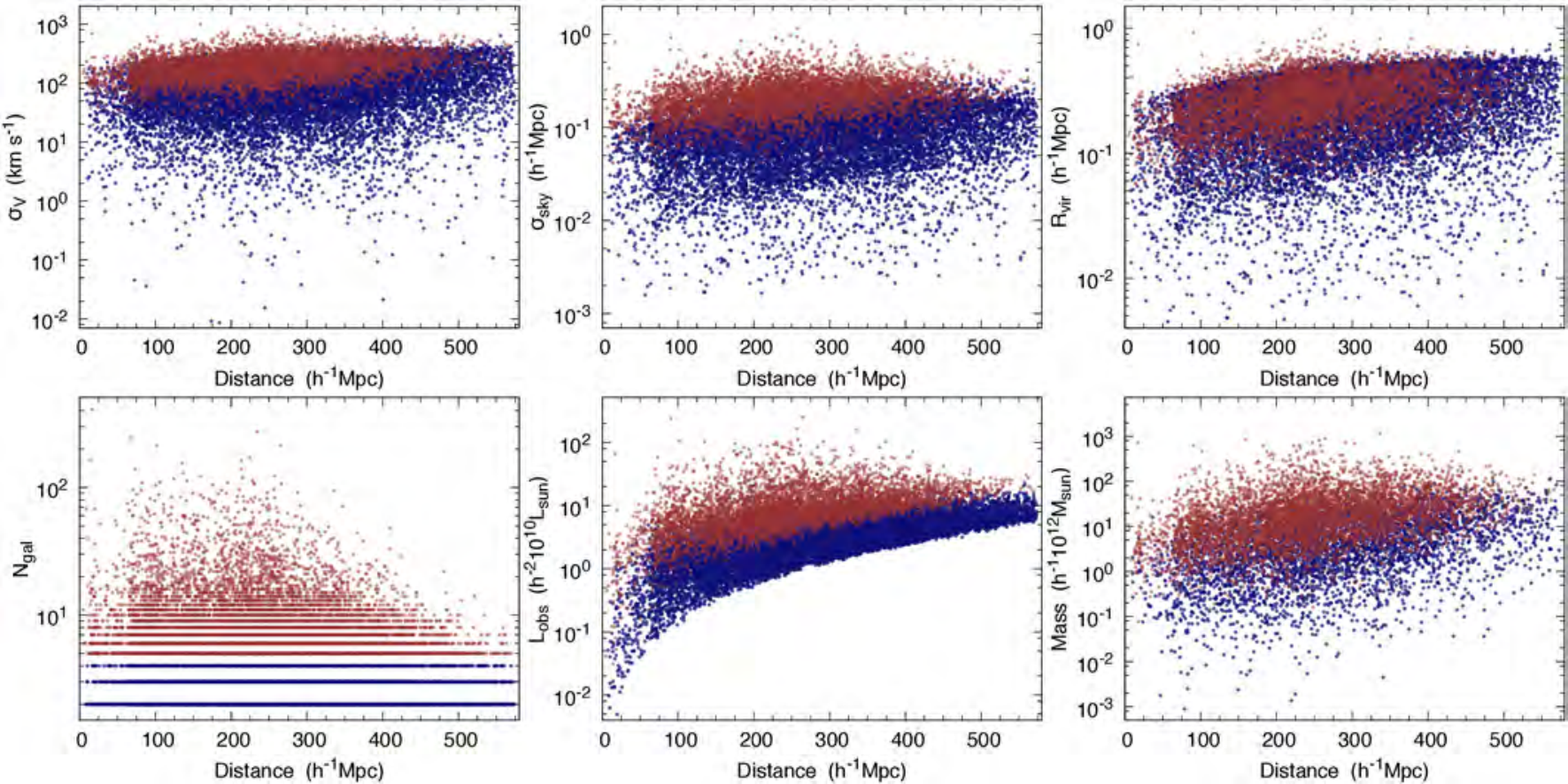}
   \caption{Flux-limited group properties as a function of comoving distance. Upper left -- velocity dispersion; upper middle -- group extent in the plane of the sky; upper right -- virial radius; lower left -- number of galaxies in the group; lower middle -- sum of the observed galaxy luminosities; lower right~-- mass of the galaxy group. Blue and red dots show groups with less than five and five or more members, respectively.}
   \label{fig:gr_dist}
\end{figure*}

%==========================================================================
\section{Dynamical masses of the groups and clusters}

To estimate the total masses of galaxy systems in the catalogue, we used the virial theorem $2T=U$, where $T$ is kinetic energy and $U$ is potential energy. Expressing

\begin{equation} 
	T=\frac{M_\mathrm{tot}\sigma_v^2}{2}, \qquad U = G\frac{M_\mathrm{tot}^2}{R_g}, \label{eq:pot}
\end{equation} 
where $\sigma_v$ is the velocity dispersion within the system, $M_\mathrm{tot}$ is the total mass of the system, $G$ is the gravitational constant, and $R_g$ is the gravitational radius, we can derive the following relation for the total mass:
\begin{equation} 
	M_\mathrm{tot} = 
	2.325\times 10^{12} 
	\frac{R_g}{\mathrm{Mpc}} 
	\left( \frac{\sigma_v}{100\,\mathrm{km\,s}^{-1}}
	\right)^2 
	M_{\sun} .
	\label{eq:mtot}
\end{equation}
Thus in order to estimate the masses of our FoF groups, we have to find two quantities: $R_g$ and $\sigma_v$. Velocity dispersion is estimated using the line-of-sight velocities of all detected galaxies of a galaxy system. The calculated quantity is the one-dimensional velocity dispersion $\sigma_{v1D}$ (Eg.~\ref{eq:sigv}). Assuming dynamical symmetry, the real (3D) velocity dispersion in groups would thus be $\sigma_v = \sqrt{3}\sigma_{v1D}$.

According to \citet{Binney:08}, $R_g$ can be found by equalising the potential energy (\ref{eq:pot}) with the equation
\begin{equation} 
	U = 4\pi G \int\limits_0^{R_\mathrm{out}}
\frac{M(r)}{r}\rho(r)r^2\mathrm{d}r, \label{eq:pot_bt} 
\end{equation} 
where $M(r)$ is the mass within a sphere $r$ and $R_\mathrm{out}$ is the outer limit of the system. The assumed density profile for the galaxy systems (groups/clusters) is $\rho(r)$.

To estimate $R_g$ from observations, we have used the observed dispersion in the plane of the sky $\sigma_\mathrm{sky}$ (Eq.~\ref{eq:sigr}). To do that, we assumed some density profile (NFW and Hernquist) and calculated the relation $\kappa = R_g/\sigma_\mathrm{sky}$ based on the assumed density profile. We used $\kappa$ to transfer the observed $\sigma_\mathrm{sky}$ to $R_g$, which was then used in Eq.~(\ref{eq:mtot}) to calculate the dynamical masses of the groups.

The derivation of the group masses for the cases of NFW and Hernquist mass distribution are described below. The masses are estimated only for groups with three or more members. For galaxy pairs, the group extent in the sky and velocity dispersion are not clearly defined. But also for other poorer groups the estimated group mass is largely uncertain.

\subsection{NFW profile}

Assuming the NFW profile \citep{Navarro:97}, the $\rho(r)$ depends only on one parameter: the halo mass
$M_{200}$, which all the other parameters depend on. The profile is expressed as
\begin{equation} 
	\rho(r) = \frac{\delta_c
\rho_\mathrm{crit}}{\left(\frac{r}{R_s}\right)\left(1+\frac{r}{R_s}\right)^2},
\end{equation}
where $\rho_\mathrm{crit}$ is the critical density of the Universe and
$R_s$ is a scale radius. Defining the concentration as $c_{200}=R_{200}/R_s$, where
$R_{200}$ is the radius containing the mass $M_{200}$, while the mean
density inside that radius is 200 times the critical density of the Universe, the $\delta_c$ is expressed
as 
\begin{equation}
	\delta_c =
\frac{200}{3}\frac{c_{200}^3}{\ln(1+c_{200})-\frac{c_{200}}{1+c_{200}}}.
\end{equation}
Following \citet{Maccio:08}, the parameter $c_{200}$ is related to the halo mass $M_{200}$ as
\begin{equation} 
	\log(c_{200}) = 0.83 - 0.098 \log \left[
\frac{M_{200}} {10^{12}h^{-1}M_{\sun}} \right] . 
\end{equation}

To calculate the ratio $\kappa$ assuming the NFW profile, we take $R_\mathrm{out}$ in
Eq.~(\ref{eq:pot_bt}) equal to $R_{200}$. This allows to calculate the quantity $\kappa$ for a fixed mass $M_\mathrm{tot}$ .

The $\sigma_\mathrm{sky}$ for the NFW profile is calculated by integrating the projected NFW density \citep{Bartelmann:96, Lokas:01}.

The total mass of the group is found iteratively. Firstly, we assume some mass
$M_\mathrm{tot}$, which is taken to be $M_{200}$ for the NFW profile. Using this
$M_{200}$, we can determine the NFW profile and compute the $\kappa$ as described above.
Using the estimated $\kappa$ value, we recalculate the total mass of the group
$M_\mathrm{tot}$. Based on this mass we recalculate $\kappa$. We iterate the process, until
it converges. Usually, it takes less than ten iterations to converge.

\begin{figure}
   \centering
   \includegraphics[width=88mm]{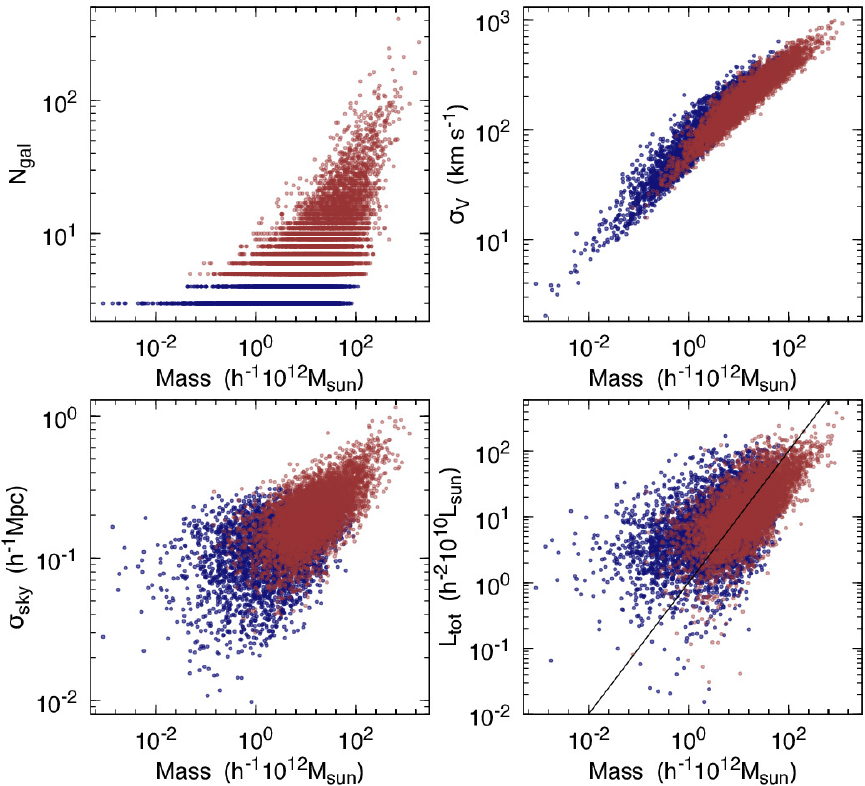}
   \caption{Flux-limited group masses versus group properties: upper left~-- group richness; upper right -- velocity dispersion; lower left -- group extent; lower right -- group total luminosity. Blue points denote small groups (less than 5 members) and red dots denote larger groups. Galaxy pairs are excluded from this figure.}
   \label{fig:gr_mass}
\end{figure}

\begin{figure}
   \centering
   \includegraphics[width=88mm]{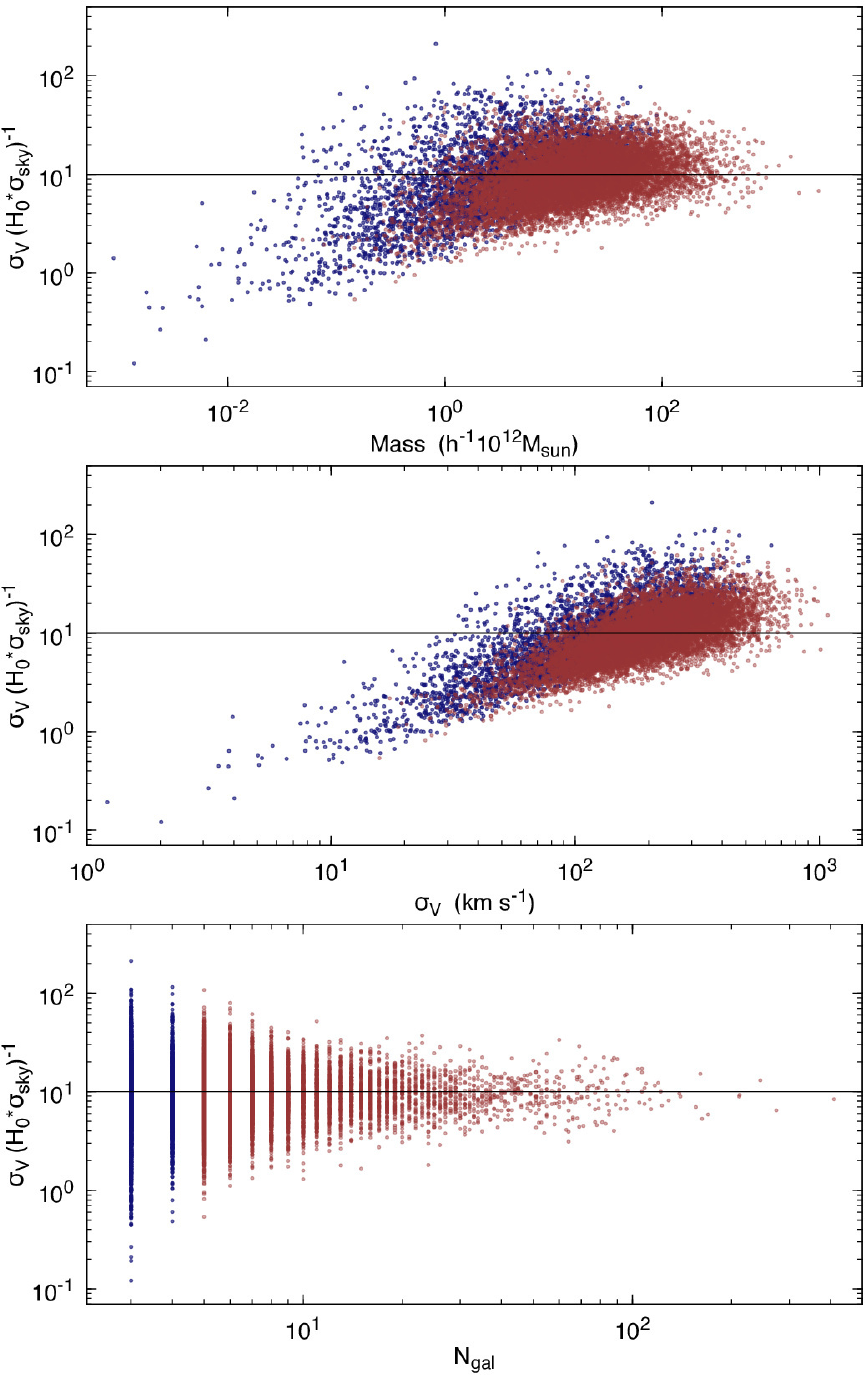}
   \caption{Group shape in the redshift space (ratio of velocity dispersion to the group extent in the sky) as a function of group mass (upper panel), velocity dispersion (middle panel), and group richness (lower panel). Blue points show poor groups (less than five members) and red dots show richer groups.}
   \label{fig:gr_shape}
\end{figure}

\begin{figure}
   \centering
   \includegraphics[width=88mm]{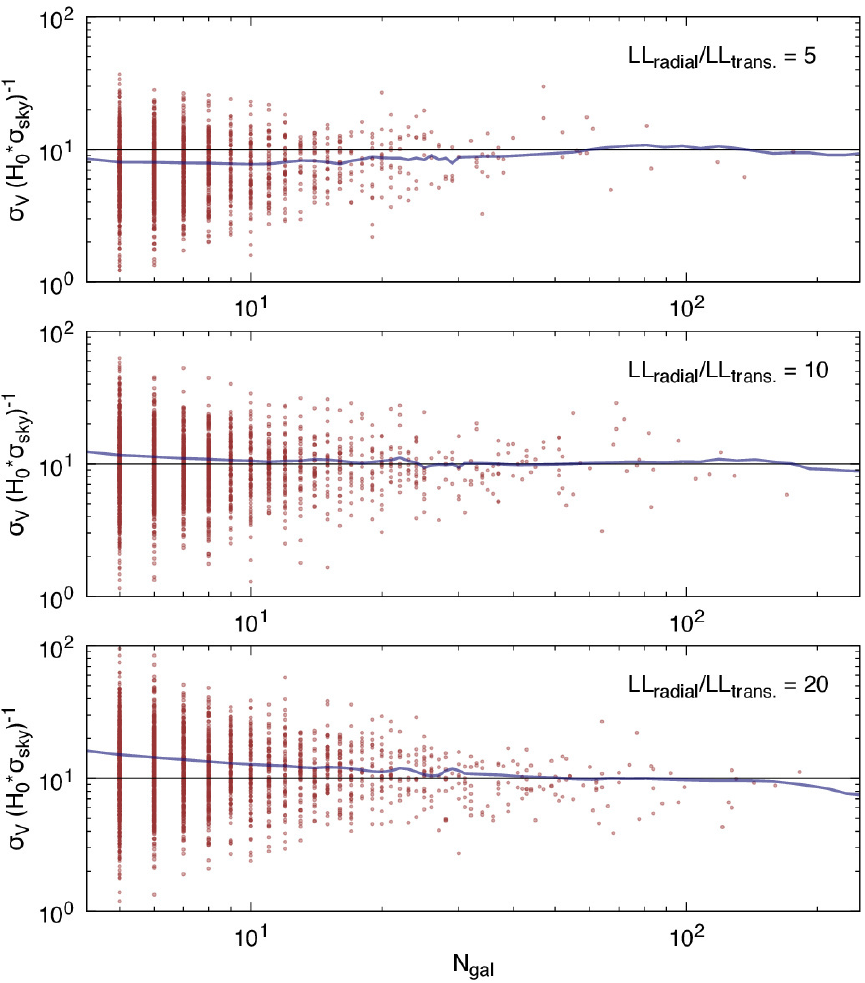}
   \caption{Group shape in the redshift space as a function of group richness for different input LL ratios. Blue solid line shows the running mean for each sample.}
   \label{fig:testgr_shape}
\end{figure}

\subsection{Hernquist profile}

The Hernquist profile \citep{Hernquist:90} is similar to the NFW profile at the centre, but has a finite mass:
\begin{equation}
\rho(r)=\frac{M}{2\pi}\frac{a}{r}\frac{1}{(r+a)^3},
\end{equation}
where $M$ is the total mass of the halo and $a$ is the scale radius. 

For this profile, the total potential energy $U$ is
\begin{equation}
U= - \frac{GM^2}{6a}      
\end{equation}
\citep[see Eq.~14 in][]{Hernquist:90}. This  gives $R_g=6a$.

The scale factor $a$ is easy to find using  the half-mass radius $R_e$ of the projected density distribution,
$R_e=1.8153 a$ \citep[see Eq.~38 in][]{Hernquist:90}.

The half-mass radius $R_e$ is the median of the sample of the projected galaxy distances from the centre of the group, and can be estimated directly from observations. This estimate is, however, noisy, especially for low-richness groups, and so we estimate it on the basis of our $\sigma_{\mathrm{sky}}$ (Eq.~\ref{eq:sigr}). This amounts to approximating the group galaxy distribution in the sky by a 2D Gaussian
\begin{equation}
f(x,y)\mathrm{d}x\mathrm{d}y=\frac{1}{2\pi\sigma_\mathrm{sky}^2}\exp(-\frac{x^2+y^2}{2\sigma_\mathrm{sky}^2})\mathrm{d}x\mathrm{d}y,
\end{equation}
where $x,y$ are any local Cartesian coordinates in the plane of the sky.

As $x^2+y^2=R^2$ and $\mathrm{d}x\mathrm{d}y=2\pi R \mathrm{d}R$ this is the Rayleigh distribution for $R$:
\begin{equation}
f(R)\mathrm{d}R=\frac{1}{\sigma_\mathrm{sky}^2}\exp(-R^2/2\sigma_\mathrm{sky}^2)R \mathrm{d}R,
\end{equation}
and 
\begin{equation}
\int_0^\infty R^2 f( R)\mathrm{d}R=2\sigma_\mathrm{sky}^2, 
\end{equation}
giving the estimate 
\begin{equation}
\hat{\sigma}^2_\mathrm{sky}=\frac1{2N}\sum_{i=1}^{N} R_i^2
\end{equation}
that we give in our catalogue.

Now, for the Rayleigh distribution the integral probability 
$P(R<R_a)=F(R_a)=1-\exp(-R_a^2/2\sigma_\mathrm{sky}^2)$. The half-mass
condition says $F(R_e)=1/2$,
giving $R_e=\sigma_\mathrm{sky}\sqrt{2\ln 2}=1.386 \sigma_\mathrm{sky}$.
Equating this with the Hernquist $R_e$ above, we get
$R_e=1.386\sigma_\mathrm{sky}=1.8153 a$, hence  $a=0.764\sigma_\mathrm{sky}$
and $R_g=6a=4.582\sigma_\mathrm{sky}$. 

The estimated masses using Hernquist and NFW profiles are tightly related. Figure~\ref{fig:gr_mass_nfw_her} shows the ratio of these two mass estimates. It is seen that Hernquist masses are 1.55--1.75 more massive, depending on the system mass. This can be also considered as the systematic bias that comes from the assumption of using a fixed mass profile.

%==========================================================================

\section{Results: group and cluster catalogues}

\subsection{Properties of flux-limited groups and clusters}

Various properties of the flux-limited groups are shown in Fig.~\ref{fig:gr_dist} as a function of distance. We see that the main observational properties of the groups, radial velocity dispersion and group extent in the sky, do not depend much on distance. A slight correlation with distance is expected since the derived group properties depend on group richness and/or the imposed magnitude limit. For example, \citet{Old:13} show that the velocity dispersion is slightly underestimated when only the brightest cluster members are considered. In general, Fig.~\ref{fig:gr_dist} shows that the flux-limited selection effect has been largely eliminated while choosing the distance-dependent LL for the FoF. 

The distribution of group richness as a function of distance (lower left panel in Fig.~\ref{fig:gr_dist}) indicates that further away, richer groups become scarce. This is a natural result for a flux-limited survey. Nevertheless, some rich groups (more than 50 members) can be found up to 400~$h^{-1}\mathrm{Mpc}$.

Figure~\ref{fig:gr_mass} shows the flux-limited group richness, velocity dispersion, group extent, and the total luminosity of the group as a function of group mass. We see that the tightest correlation is between the group mass and the velocity dispersion. This is expected since the virial theorem is most sensitive to velocity dispersion. The upper-left panel in this figure also shows that for small groups the scatter of mass estimation is very large. The scatter decreases while moving toward richer groups.

The relation between the group mass and the total luminosity of the group is plotted in the lower-right panel in Fig.~\ref{fig:gr_mass}. The total luminosity given by Eq.~(\ref{eq:totlum}) is compensated for the luminosity of galaxies which are missed due to the magnitude limit of the survey. This estimate of the total luminosity is a statistically better characteriser of groups than the total observed luminosity or the group richness. However, it is not accurate for any particular group since the luminosity function depends on group properties (e.g. mass). The scatter in this plot is relatively large, however, the correlation is present. Since the total luminosity can be used to estimate the environmental density (using the smoothed luminosity density fields), this plot hints that the restored luminosity density field is a realistic characteriser of the environment.

\begin{table}
\caption{Volume-limited group statistics. Comparison with flux-limited groups.}
\label{tab:vollim-stat}
\centering
\begin{tabular}{lccc}
\hline\hline
Sample & Frac\tablefootmark{a} & Frac 0.5\tablefootmark{b} & Frac 0.1\tablefootmark{c} \\
 & \% & \% & \% \\
\hline
Volume-limited-18.0 & 91 & 71 & 37 \\
Volume-limited-18.5 & 90 & 71 & 38 \\
Volume-limited-19.0 & 89 & 70 & 39 \\
Volume-limited-19.5 & 87 & 69 & 40 \\
Volume-limited-20.0 & 83 & 63 & 37 \\
Volume-limited-20.5 & 80 & 60 & 39 \\
Volume-limited-21.0 & 76 & 56 & 39 \\
\hline
\end{tabular}
\tablefoot{
\tablefoottext{a}{Match fraction between volume-limited groups and flux-limited groups. Match radius is 1~$h^{-1}\mathrm{Mpc}$.}
\tablefoottext{b}{Fraction of matches where relative mass difference is smaller than 0.5.}
\tablefoottext{c}{Fraction of matches where relative mass difference is smaller than 0.1.}
}
\end{table}

Figure~\ref{fig:gr_shape} indicates the group shape in the redshift space (the ratio of the velocity dispersion to the group extent in the plane of sky) as a function of the group mass, velocity dispersion, and group richness. We see that the group shape does not depend on the group richness and there is a small dependency on the group mass. However, the group shape has a clear dependency on the group velocity dispersion. The groups/clusters that have large velocity dispersion are more stretched out in the redshift space. The same dependency was recently shown by \citet{Wojtak:13}.

In Fig.~\ref{fig:gr_shape} the average elongation of groups is roughly 10. The same value has been used for the ratio between LLs in radial and transversal directions. One might think that the derived average elongation of groups depends solely on the input LL ratios. To test this, we generated two test group catalogues, where the input LL ratio was 5 and 20, respectively, while all the other parameters were kept the same. The resultant average elongation of groups is shown in Fig.~\ref{fig:testgr_shape}. We see that the average elongation in each case is still roughly 10; only the smallest groups display a slight dependency on the input LL ratio. We conclude that the derived elongation of groups (except for poor groups) does not depend significantly on the input LL ratio.

\subsection{Volume-limited groups and mass function}
\label{sect:res_vollim}

\begin{figure*}
   \centering
   \includegraphics[width=180mm]{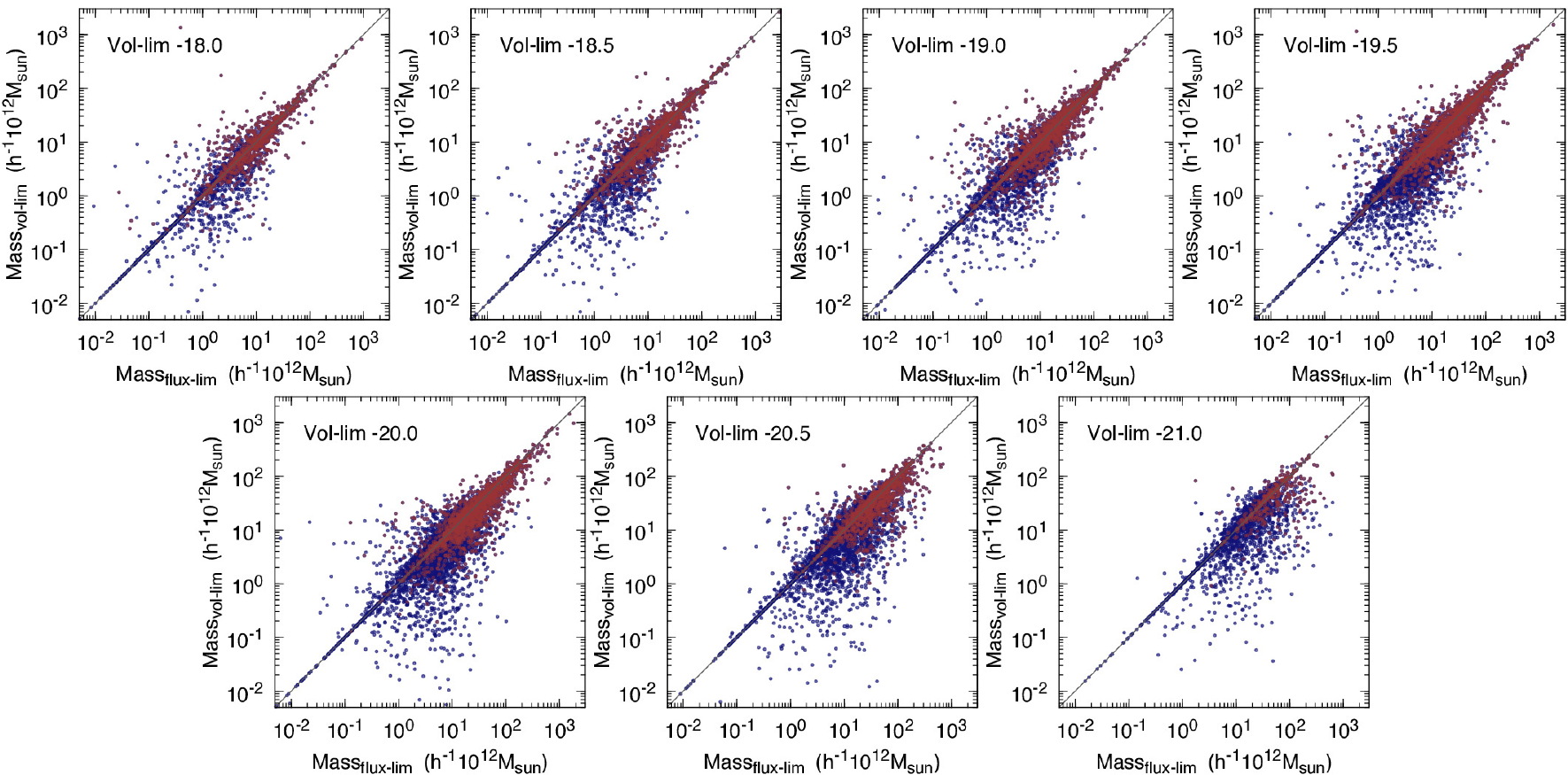}
   \caption{Flux-limited group/cluster masses compared to the volume-limited group/cluster masses. Red points show the groups with five or more members and blue points show poorer groups. The solid grey line shows the one-to-one relationship between flux- and volume-limited masses.}
   \label{fig:mass_comp}
\end{figure*}

Table~\ref{tab:vollim-stat} lists the fraction of the volume-limited groups that are present also in the flux-limited sample. Matches between the volume- and flux-limited groups have been found within a 1~$h^{-1}\mathrm{Mpc}$. We see that overall, the fraction of matches is 80--90\%, yielding that most of the volume-limited groups are identified at the same locations as the flux-limited groups. Table~\ref{tab:vollim-stat} also gives the fraction of groups in the volume-limited samples that have roughly the same mass as estimated for the corresponding groups in the flux-limited sample. Up to the relative mass difference 0.5, the fraction of matches is 60--70\%. Even within the relative mass difference 0.1, the fraction of matches is roughly 40\%. Considering the large scatter of the mass estimations (especially for poorer groups), the fact that majority of the groups has relatively close mass estimations indicates the reliability of the mass determination method and also means that the groups identified using different samples are the same physical objects. We conclude that the used strategy for the LLs in the volume-limited samples is appropriate and the volume-limited groups are well compatible with the flux-limited groups.

Figure~\ref{fig:mass_comp} shows a comparison of the masses of the volume-limited groups and the masses of the corresponding flux-limited groups. We see that for richer groups the masses in the flux- and volume-limited samples are rather well correlated, indicating that the groups from different samples are generally overlapping. For poorer groups, the masses in the volume-limited samples tend to be underestimated due to a too high fraction of group members being missed because of the flux limit.

\begin{figure}
   \centering
   \includegraphics[width=88mm]{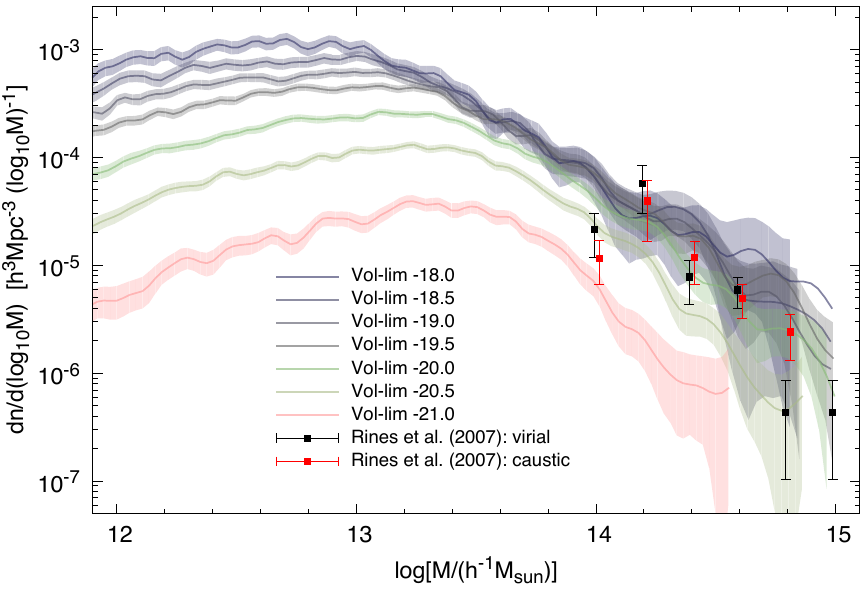}
   \caption{Group mass function for various volume-limited samples (solid lines) together with 95\% confidence limits (shaded regions). The mass function is calculated using groups with three or more members. Black and red points and error bars are mass function from \citet{Rines:07} using virial masses and caustic masses, respectively (the error bars show 68\% uncertainties).}
   \label{fig:gr_massfun}
\end{figure}

The correspondence between the flux- and volume-limited samples can also be checked by comparing the resultant group mass functions, as shown in Fig.~\ref{fig:gr_massfun}. We see that the mass function for massive clusters in different samples is similar (except for the brightest volume-limited sample), thus the masses estimated using only the brightest cluster galaxies are as reliable as the masses estimated using also fainter galaxies. The mass function of groups in the brightest volume-limited sample with $M_{r,\mathrm{lim}} =-21.0$ lies notably below the other samples. In this case, too few galaxies remain within the luminosity limits and group sample is incomplete even for massive clusters. In several cases, only 1--2 brightest galaxies of each group/cluster remain within the luminosity limits and these groups/clusters are not present in Fig.~\ref{fig:gr_massfun}. At lower masses, the samples start to progressively deviate from each other because the number of detected systems depends on the brightness limit.

For comparison, the cluster mass functions derived by \citet{Rines:07} are plotted in Fig.~\ref{fig:gr_massfun}. They studied the cluster masses in $X$-ray-selected sample of clusters with data from SDSS DR4. The masses were computed in two ways, one using the virial theorem (black squares) and the other (red squares) using the caustic technique \citep{Rines:06}. Both mass functions are close to our mass functions for massive clusters. \citet{Rines:07} compared their results also with the mass functions estimated with other methods: using $X$-ray data \citep{Reiprich:02} and using the mass-richness relation, applied on early SDSS data \citep{Bahcall:03}. All of these independent methods agree well in the massive clusters domain.

Figure~\ref{fig:gr_massfun} shows that at the low end, the mass function depends on the applied magnitude limit. The higher the absolute magnitude limit, the smaller is the number of detected groups and clusters. The slope of the mass function for different volume-limited samples is similar, which indicates that the group/cluster detection algorithm does not create mass-dependent systematic errors. In the near future, we plan to study the mass function of groups and clusters in the SDSS data and in different cosmological models in more detail.

%==========================================================================

\section{Conclusions}

We have updated and improved our previous galaxy group catalogues, constructed on the basis of the FoF method, following a similar procedure as used in the previous papers \citep{Tago:08,Tago:10,Tempel:12a}. The group finding method is applied to flux- and volume-limited samples drawn from the SDSS main contiguous area, covering 7221 square degrees in the sky. In addition to the SDSS spectroscopic redshifts, the galaxy sample is complemented with redshifts from the 2MRS, 2dFGRS and RC3 catalogues.

As an important addition, we have estimated the group/cluster dynamical masses using the virial theorem. We have shown that the groups extracted from the flux- and volume-limited samples are well compatible with each other. We have calculated the mass functions for the volume-limited samples. At the massive end, the functions are in agreement with previous observational estimates \citep{Rines:07}.

As a next step, we plan to use the new catalogue to map the large-scale galactic filaments \citep[similarly to][]{Tempel:14} and to construct a catalogue of superclusters \citep[similarly to][]{Liivamagi:12}.
%The study of various structure elements together will help to improve our knowledge about the evolution of the large-scale structure of the Universe and to understand the connection between different structure elements.

%==========================================================================
\begin{acknowledgements}
	
	We thank our colleagues from Tartu Observatory and Tuorla Observatory for carefully checking the catalogues. We acknowledge the support by the Estonian Research Council grants 9428, MJD272, PUT246, IUT26 and the Centre of Excellence of Dark Matter in (Astro)particle Physics and Cosmology. All the figures have been made using the gnuplot plotting utility. This research has made use of the NASA’s Astrophysics Data System Bibliographic Services.
	
	Funding for SDSS-III has been provided by the Alfred P. Sloan Foundation, the Participating Institutions, the National Science Foundation, and the U.S. Department of Energy Office of Science. The SDSS-III web site is \url{http://www.sdss3.org/}.

	  SDSS-III is managed by the Astrophysical Research Consortium for the Participating Institutions of the SDSS-III Collaboration including the University of Arizona, the Brazilian Participation Group, Brookhaven National Laboratory, Carnegie Mellon University, University of Florida, the French Participation Group, the German Participation Group, Harvard University, the Instituto de Astrofisica de Canarias, the Michigan State/Notre Dame/JINA Participation Group, Johns Hopkins University, Lawrence Berkeley National Laboratory, Max Planck Institute for Astrophysics, Max Planck Institute for Extraterrestrial Physics, New Mexico State University, New York University, Ohio State University, Pennsylvania State University, University of Portsmouth, Princeton University, the Spanish Participation Group, University of Tokyo, University of Utah, Vanderbilt University, University of Virginia, University of Washington, and Yale University.
\end{acknowledgements}

%\bibliographystyle{aa}
%\bibliography{mybib}{}

%==========================================================================
\appendix

\section{The catalogues}
\label{app:cat}

    The catalogue of flux- and volume-limited groups and clusters of galaxies consists of several tables: two tables for each sample. The first table lists the galaxies that were used to generate the catalogue of groups and clusters, and the second one describes the group properties. The flux-limited galaxy table includes all the relevant parameters for galaxies. The volume-limited galaxy table includes only the basic parameters and references to the flux-limited sample. The structure of group catalogues for the flux- and volume-limited samples is identical.
    
The catalogues are available at \url{http://cosmodb.to.ee}.
The catalogues will be made available also through the Strasbourg Astronomical Data Centre (CDS).

\subsection{Description of the flux-limited galaxy catalogue}

The flux-limited galaxy catalogue contains the following information (column numbers are given in square brackets):
\begin{enumerate}
 \item{[1]\,\texttt{galid} --} our unique identification number for galaxies;
 \item{[2]\,\texttt{specobjid} --} SDSS DR10 spectroscopic object identification number;
 \item{[3]\,\texttt{objid} --} SDSS DR10 photometric object identification number;
 \item{[4]\,\texttt{groupid} --}  group/cluster id;
 \item{[5]\,\texttt{ngal} --} richness (number of members) of the group/cluster the galaxy belongs to;
 \item{[6]\,\texttt{rank} --} luminosity rank of the galaxy within its group; rank~1 indicates the most luminous galaxy;
 \item{[7]\,\texttt{groupdist} --} comoving distance to the group/cluster centre, where the galaxy belongs to, in units of $h^{-1}\,$Mpc, calculated as an average over all galaxies within the group/cluster;
 \item{[8]\,\texttt{zobs} --} observed redshift (without the CMB correction), as given in the SDSS CAS;
 \item{[9]\,\texttt{zcmb} --} redshift, corrected to the CMB rest frame;
 \item{[10]\,\texttt{zerr}--} uncertainty of the redshift;
 \item{[11]\,\texttt{dist} --} comoving distance in units of $h^{-1}\,$Mpc (calculated directly from the CMB-corrected redshift);
 \item{[12]\,\texttt{dist\_cor} --} comoving distance of the galaxy after suppressing of the finger-of-god effect (as used in the luminosity density field calculations);
 \item{[13--14]\,\texttt{raj2000, dej2000} --} right ascension and declination (deg);
 \item{[15--16]\,\texttt{glon, glat} --} galactic longitude and latitude (deg);
 \item{[17--18]\,\texttt{sglon, sglat} --} supergalactic longitude and latitude (deg);
 \item{[19--20]\,\texttt{lam, eta} --} SDSS survey coordinates $\lambda$ and $\eta$ (deg);
 \item{[21--3]\,\texttt{crd\_xyz} --} Cartesian coordinates defined by $\eta$ and $\lambda$;
 \item{[24--28]\,\texttt{mag\_$x$} --} Galactic-extinction-corrected Petrosian magnitude ($x\in ugriz$ filters);
 \item{[29--33]\,\texttt{absmag\_$x$} --} absolute magnitude of the galaxy, $k$+$e$-corrected ($x\in ugriz$ filters, in units of $\mathrm{mag}+5\log_{10}h$);
 \item{[34--38]\,\texttt{kecor\_$x$} --} $k+e$-correction ($x\in ugriz$ filters);
 \item{[39--43]\,\texttt{ext\_$x$}} Galactic extinction ($x\in ugriz$ filters);
 \item{[44]\,\texttt{lum\_r} --} observed luminosity in the $r$-band in units of $10^{10}h^{-2}L_\odot$, where $M_\odot=4.64$ \citep{Blanton:07};
 \item{[45]\,\texttt{weight} --} weight factor for the galaxy (\texttt{w}$\cdot$\texttt{lum\_r} was used to calculate the luminosity density field);
 \item{[46]\,\texttt{source} --} source of the redshift: 0 for SDSS, 1 for 2MRS, 2 for 2dFGRS, 3 for RC3;
 \item{[47]\,\texttt{bad\_lum} --} if set to 1, galaxy luminosity is uncertain;
 \item{[48]\,\texttt{morf} --} morphology of the galaxy (0 -- unclear, 1 -- spiral, 2 -- elliptical);
 \item{[49]\,\texttt{morf\_zoo} --} morphology from the galaxy zoo project \citep{Lintott:08}: 0 -- unclear, 1 -- spiral, 2 -- elliptical;
 \item{[50]\,\texttt{hc\_e} --} probability of being an early-type galaxy \citep[from][]{Huertas-Company:11};
 \item{[51]\,\texttt{hc\_s0} --} probability of being an S0 galaxy;
 \item{[52]\,\texttt{hc\_sab} --} probability of being an Sab galaxy;
 \item{[53]\,\texttt{hc\_scd} --} probability of being an Scd galaxy;
 \item{[54]\,\texttt{dist\_edge} --} comoving distance of the galaxy from the border of the survey mask;
 \item{[55--58]\,\texttt{den$a$} --} normalised environmental density of the galaxy for various smoothing scales ($a=1,\,2,\,4,\,8$~$h^{-1}\,$Mpc).
\end{enumerate}

\subsection{Description of the volume-limited galaxy catalogue}

The volume-limited galaxy catalogues contain the following information (column numbers are given in square brackets):
\begin{enumerate}
  \item{[1]\,\texttt{galid} --} our unique identification number for galaxies (identical with the flux-limited galaxy identification number);
  \item{[2]\,\texttt{groupid} --}  group/cluster id, unique within one volume-limited sample;
  \item{[3]\,\texttt{ngal} --} richness (number of members) of the group the galaxy belongs to;
  \item{[4]\,\texttt{rank} --} luminosity rank of the galaxy within its group; rank~1 indicates the most luminous galaxy;
  \item{[5--9]\,\texttt{absmag\_$x$} --} absolute magnitude of the galaxy, $k$+$e$-corrected ($x\in ugriz$ filters, in units of $\mathrm{mag}+5\log_{10}h$);
  \item{[10]\,\texttt{lum\_r} --} observed luminosity in the $r$-band in units of $10^{10}h^{-2}L_\odot$, where $M_\odot=4.64$ \citep{Blanton:07};
  \item{[11]\,\texttt{dist\_cor} --} comoving distance of the galaxy after suppressing the finger-of-god effect;
  \item{[12--14]\,\texttt{crd\_xyz} --} Cartesian coordinates defined by $\eta$ and $\lambda$.
\end{enumerate}

\subsection{Description of the group/cluster catalogues}

The catalogue of groups/clusters contains the following information (column numbers are given in square brackets):
\begin{enumerate}
    \item{[1]\,\texttt{groupid} --} group/cluster id;
    \item{[2]\,\texttt{ngal} --} richness (number of members) of the group;
    \item{[3--4]\,\texttt{raj2000, dej2000} --} right ascension and declination of the group centre (deg);
    \item{[5]\,\texttt{zcmb} --} CMB-corrected redshift of group, calculated as an average over all group/cluster members;
    \item{[6]\,\texttt{groupdist} --} comoving distance to the group centre ($h^{-1}$Mpc);
    \item{[7]\,\texttt{sigma\_v} --} rms radial velocity deviation ($\sigma_V$ in physical coordinates,  in \mbox{km\,s$^{-1}$});
    \item{[8]\,\texttt{sigma\_sky} --} rms deviation of the projected distance in the sky from the group centre ($\sigma_\mathrm{sky}$ in physical coordinates, in $h^{-1}$Mpc), $\sigma_\mathrm{sky}$ defines the extent of the group in the sky;
    \item{[9]\,\texttt{r\_vir} --} virial radius in $h^{-1}$Mpc (the projected harmonic mean, in physical coordinates);
    \item{[10]\,\texttt{r\_max} --} maximum radius of the group/cluster;
    \item{[11]\,\texttt{mass\_nfw} --} estimated mass of the group assuming the NFW density profile (in units of $10^{12}h^{-1}M_\odot$);
    \item{[12]\,\texttt{mass\_her} --} estimated mass of the group assuming the Hernquist density profile (in units of $10^{12}h^{-1}M_\odot$);
    \item{[13]\,\texttt{lum\_r\_group} --} observed luminosity, i.e. the sum of the luminosities of the galaxies in the group/cluster ($10^{10}h^{-2}L_\odot$);
    \item{[14]\,\texttt{weight} --} weight factor for the group at the mean distance of the group;
    \item{[15--18]\,\texttt{den$a$} --} normalised environmental density (mean of group galaxy densities) of the group for various smoothing scales ($a=1,\,2,\,4,\,8$~$h^{-1}\,$Mpc).
\end{enumerate}    

\subsection{Access to the SDSS database}

To facilitate the use of all the parameters available in the SDSS CAS, we have uploaded our galaxy \texttt{galid}-s together with the SDSS \texttt{objid} and \texttt{specobjid} to the CAS server. For example, to add a Galactic-extinction-corrected model magnitude in the $r$-band (\texttt{dered\_r}) from the SDSS CAS to our catalogue, the following SQL query in CAS can be used:
\begin{verbatim}
select temp.galid, dered_r 
  from DR10.PhotoObjAll as ph
  join public.elmo.Tempel_DR10 as temp
  on ph.objid = temp.objid
  into MyDB.test_db	
\end{verbatim}

%==========================================================================
\section{Estimation of galaxy luminosity evolution}
\label{app:lum_evol}

\begin{figure}
   \centering
   \includegraphics[width=88mm]{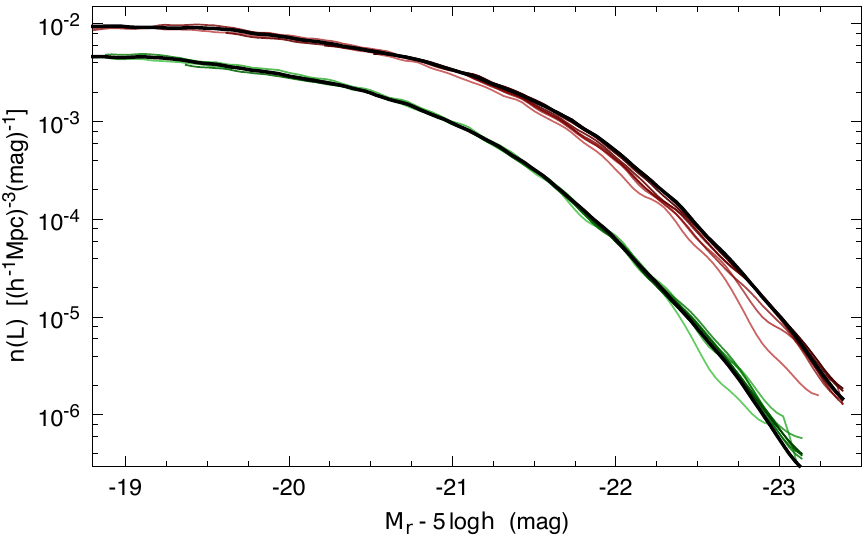}
   \caption{Upper lines show the luminosity function for various distance intervals (see text for details) without evolution corrections. Lower lines show the luminosity function with evolution correction.}
   \label{fig:lum_evol}
\end{figure}

Galaxy luminosity is not constant over time. Due to the secular evolution of galaxies, their luminosity is slightly increasing over time, even if no major influence is affecting their evolution. For a statistical analysis this small luminosity evolution has to be taken into account.

The luminosity evolution of the SDSS galaxies is estimated in \citet{Blanton:03}, where they basically assume that the luminosity function of galaxies is constant with distance. In practice, the luminosity evolution is different for various types of galaxies. For simplicity, \citet{Blanton:03} assume that the luminosity evolution for the SDSS main sample galaxies (a relatively nearby region) can be described using the equation $e_\mathrm{cor} = Qz$, where $Q$ is some constant and $z$ is redshift (Eq.~(\ref{eq:absmag}) shows how the correction is applied). In \citet{Blanton:03} the constant for the $r$-filter is estimated to be $-1.62$. The estimation of this constant is based on the early data release of the SDSS.

Below, we use the same simple approach and re-estimate the luminosity evolution of galaxies using the latest dataset, the SDSS DR10. To estimate the luminosity evolution, we calculate the luminosity function in eight distance intervals centred at 125, 175, 225, 275, 325, 375, 425, and 475~$h^{-1}\mathrm{Mpc}$: the used distance regions are $\pm 75~h^{-1}\mathrm{Mpc}$.  By changing the luminosity evolution parameter $Q$, we find the value that produces most similar luminosity functions at different distance intervals. For that we calculate the sum of the relative differences between the luminosity function in one distance interval and the luminosity function measured using all galaxies. Figure~\ref{fig:lum_evol} shows the luminosity function for the $r$ filter for two cases. The upper set of lines shows the luminosity function without the evolution correction, the lower set of lines shows the luminosity functions in various distance intervals with evolution corrections. We see that after the correction, the luminosity functions in various distance intervals are roughly the same. Our estimated evolution corrections for the $r$-filter is $-1.10$, which is slightly smaller than given by \citet{Blanton:03}.

%==========================================================================
\section{Density fields}
\label{app:den}

\begin{figure}
   \centering
   \includegraphics[width=88mm]{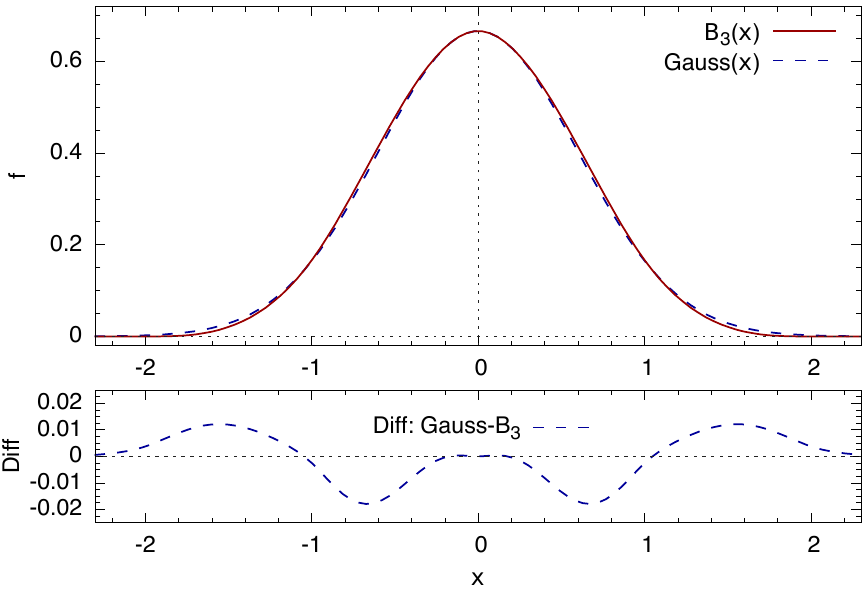}
   \caption{Gaussian (dashed line) and $B_3$ spline (solid line) kernels. Here, the $B_3$ kernel has $a=1.0$ and the Gaussian kernel has $\sigma=0.5984$. The Gaussian standard deviation $\sigma$ is chosen to have the same central density as the $B_3$ kernel.  Lower panel shows the difference between the two kernels, scaled to the central density.}
   \label{fig:kernels}
\end{figure}

\begin{figure}
   \centering
   \includegraphics[width=88mm]{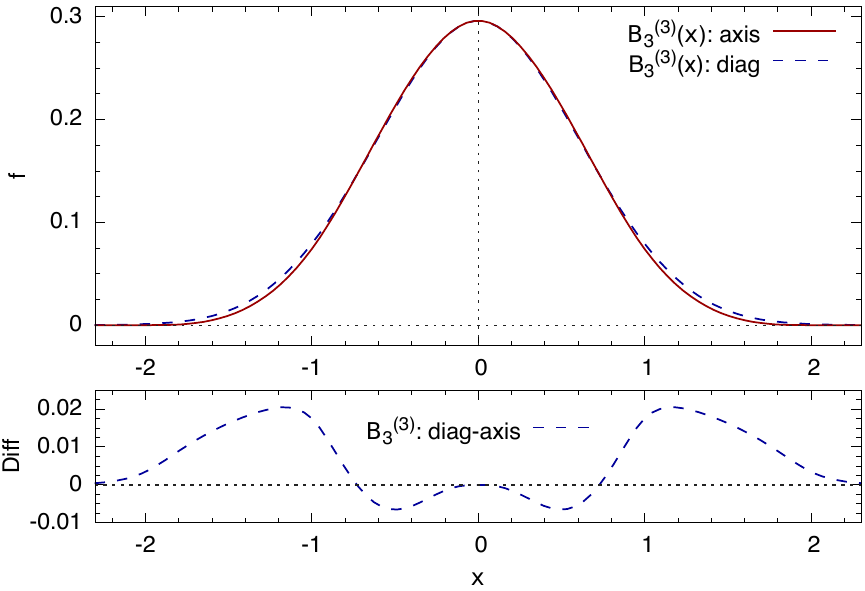}
   \caption{Deviation from isotropy for the 3D $B_3$ spline kernel. Upper panel shows $B_3$ distribution for two extreme cases: along the coordinate axis and along the diagonal of the box. Lower panel shows the difference (in units of the central density) between these extremes. The maximum deviation from isotropy is less than two per cent.}
   \label{fig:b3_asym}
\end{figure}

\begin{figure}
   \centering
   \includegraphics[width=88mm]{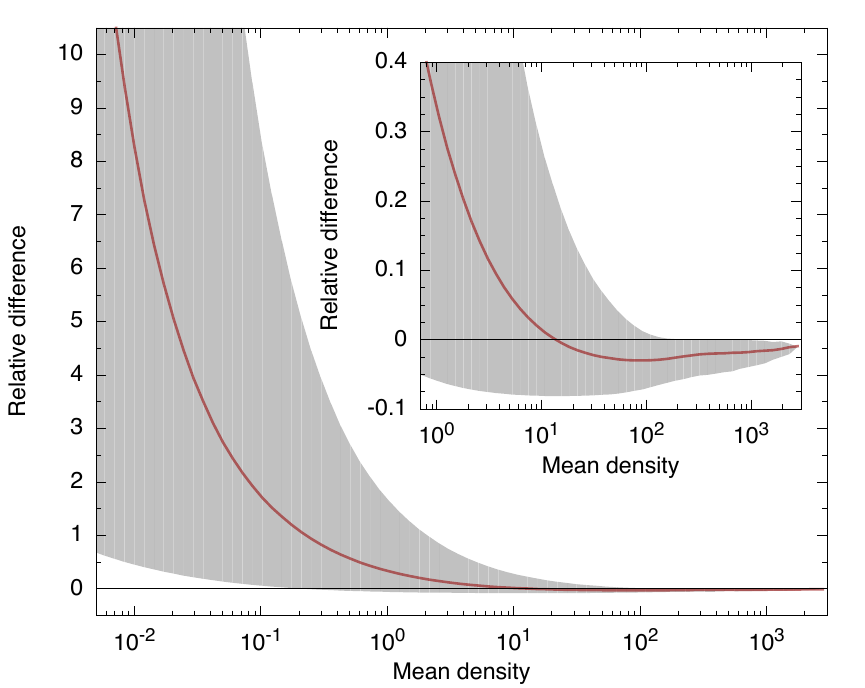}
   \caption{Relative difference ($(\ell_\mathrm{Gauss}-\ell_{B_3})/\ell_\mathrm{Gauss}$) between luminosity density fields smoothed with the Gaussian kernel and with the $B_3$ spline kernel. Density is scaled to the mean density.}
   \label{fig:b3gauss_dif}
\end{figure}

\begin{figure}
   \centering
   \includegraphics[width=88mm]{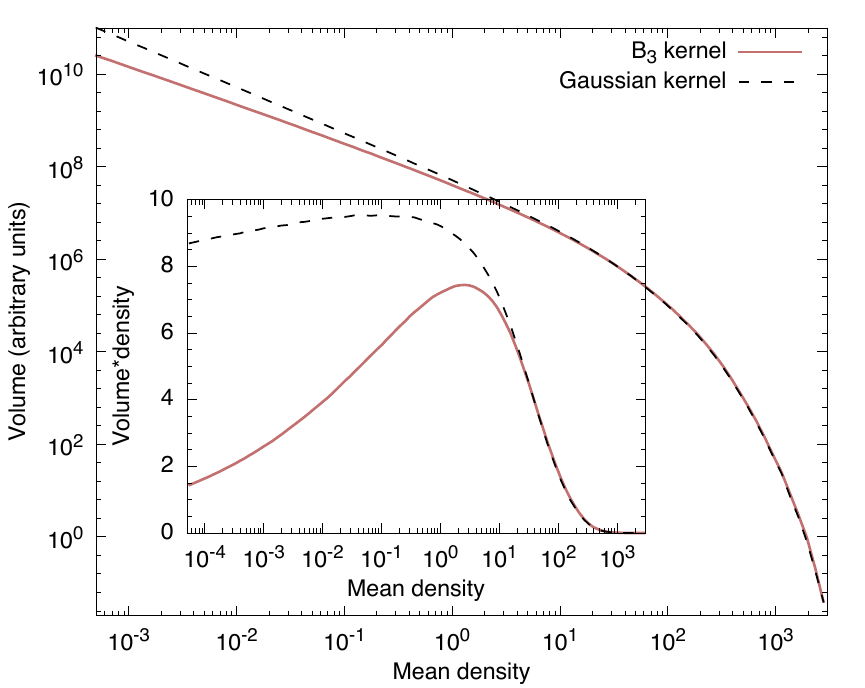}
   \caption{Number of cells for a fixed density interval. Outer figure shows the number of cells for linear density bins and inner figure for logarithmic density bins. Density is scaled to the mean density.}
   \label{fig:b3gauss_vol}
\end{figure}

	To estimate the environmental densities for galaxies and groups, we calculate the smoothed luminosity density field. Details of the calculation of the density fields are given in \citet{Liivamagi:12}. For consistency reasons, a brief description is given below.
	
	    Assuming that beside the observed galaxies, each group may also contain galaxies that lie outside of the observational window of the survey, the estimated total luminosity per one visible galaxy is
	    \begin{equation}
	        L_\mathrm{tot}=L_\mathrm{obs}\cdot W_d,
			\label{eq:totlum}
	    \end{equation}
	    where $L_\mathrm{obs}$ is the observed luminosity of the galaxy. Thus the luminosity $L_\mathrm{tot}$ takes into account the expected luminosities of the unobserved galaxies. It can, for example, be used for calculating the full luminosity density field. However, it cannot be used to estimate the total luminosity of any particular group since the luminosity function of galaxies depends on group properties.

	The distance-dependent weight factor $W_d$ is defined as
	    \begin{equation}
	        W_d = \frac{\int_0^\infty L n(L)\mathrm{d}L}{\int_{L_1}^{L_2} L n(L)\mathrm{d}L},
	    \end{equation}
	    where $L_{1,2}=L_\odot 10^{0.4(M_\odot - M_{1,2})}$ are the luminosity limits of the observational window at the distance $d$ and $n(L)$ is luminosity function. The weight factor as a function of distance is given in the inner panel of Fig.~\ref{fig:dist_vs_flux}.

    To determine the luminosity density field, we use a kernel sum:
\begin{equation}
    \ell_{\mathbf{i}} = \frac1{a^3}\sum_\mathrm{gal}  K^{(3)}\left(\frac{\mathbf{r}_\mathrm{gal} -
    \mathbf{r}_\mathbf{i}}{a}\right) L_\mathrm{tot},
    \label{eq:dfield}
\end{equation}
    where $L_\mathrm{tot}$ is the weighted galaxy luminosity and $a$ -- the kernel scale. For the kernel $K(\cdot)$ we use the $B_3$ spline function:
\begin{equation}
    \label{eq:b3}
    B_3(x) = \frac{|x-2|^3 - 4|x-1|^3 + 6|x|^3 - 4|x+1|^3 + |x+2|^3}{12}.
\end{equation}

    The luminosity density field is calculated on a regular cartesian grid generated by using the SDSS $\eta$ and $\lambda$ angular coordinates. This allows the most efficient placing of the cone inside a brick.
	
	    While calculating the density field, we also suppress the finger-of-god redshift distortions using the rms sizes of galaxy groups in the sky, $\sigma_\mathrm{sky}$, and their rms radial velocities, $\sigma_v$, (both in physical coordinates at the location of the group).  For that, we calculate the new radial distances for galaxies ($d_{\mbox{gal}}$) as
	\begin{equation}
	    d_{\mbox{gal}}=d_{\mbox{group}}+\left(d^{\star}_{\mbox{gal}}-d_{\mbox{group}}\right)\frac{\sigma_\mathrm{sky}}{\sigma_v/H_0},
	    \label{eq:distcor}
	\end{equation}
	    where $d^{\star}_{\mbox{gal}}$ is the initial distance to the galaxy, and $d_{\mbox{group}}$ is the distance to the group centre. For double galaxies, where the extent of the system in the plane of the sky does not have to show its real size (because of projection effects), we demand that its (comoving) size along the line-of-sight does not exceed the comoving LL, $d_{LL}(z)$, used to define the system:
	\begin{eqnarray}
	    d_{\mbox{gal}}&=&d_{\mbox{group}}+\left(d^{\star}_{\mbox{gal}}-d_{\mbox{group}}\right)\frac{d_{LL}(z)}{|v_1-v_2|/H_0},\\
	    &&\mbox{if}\;|v_1-v_2|/H_0 > d_{LL}(z).\nonumber
	\end{eqnarray}
	    Here $z$ is the mean redshift of the double system. If the velocity difference is smaller than that quoted above, we do not change galaxy distances.

Let us analyse the difference between the $B_3$ spline kernel and the Gaussian kernel. The kernels are plotted on Fig.~\ref{fig:kernels}, showing that the general shape of these kernels is very similar, however, $B_3$ spline kernel drops to zero behind 2.0, thus it does not extend to the infinity. In addition, the $B_3$ spline kernel is the most compact set of polynomials for a given degree, and they are interpolating, meaning that their sum over a grid is always unity. This aspect makes them ideal smoothing kernels.

Eq.~(\ref{eq:b3}) implies that the three-dimensional $B_3$ kernel is not perfectly isotropic. The deviation for extreme cases is shown in Fig.~\ref{fig:b3_asym}. It is seen that the maximum deviation is less than 2\% of the central density. Within most of the volume covered by $B_3$, the deviation remains far below 1\%. Thus the anisotropy has a negligible effect when using the kernel for density field calculations.

Next, we compare the density fields generated using the $B_3$ kernel and the Gaussian kernel. For comparison, we have fixed $a=1.0~h^{-1}\mathrm{Mpc}$ for the $B_3$ kernel and $\sigma=0.5984$ for the Gaussian kernel, yielding the same central density. Figure~\ref{fig:b3gauss_dif} shows the relative difference between the resultant density fields. It is seen that in high density regions (more than 10 times the average density), the differences are rather small. However, in low-density regions, the differences can be very high. This is expected, since Gaussian kernels extends to infinity and the density field of low-density regions becomes affected by galaxies located also in denser environments.

Figure~\ref{fig:b3gauss_vol} gives the number of cells for a fixed density interval for linear (outer panel) and logarithmic (inner panel) density bins. It is seen that the volume covered by high density regions (more than 10 times the average density) is practically the same. For lower density environments, the volume for Gaussian smoothing is actually a few times larger than for $B_3$ smoothing, distorting the density distribution in low-density environments.

To summarise, the comparison of the Gaussian and $B_3$ kernels indicates that for high density environments, the results are practically the same, while for lower density environments, the differences become rather large. Using the Gaussian smoothing, the densities in low-density environments are much higher than for the $B_3$ kernels. \citet{Saar:09} has also shown that the Minkowski functionals are rather noisy and not trustworthy in low-density regions.

\end{document}